\documentclass[usedcolumn]{raa}

\usepackage{graphicx,times}
\usepackage{natbib}
\usepackage{amssymb,amsmath}

\usepackage[a4paper=true,dvipdfm=true,pagebackref=true]{hyperref}
\hypersetup{colorlinks = true, linkcolor = green, anchorcolor = red, citecolor = blue, filecolor = red, pagecolor = red, urlcolor = red}

\begin{document}

   \title{Effects of resistivity on standing shocks in low angular momentum flows around black holes}

   \volnopage{Vol.0 (20xx) No.0, 000--000}      
   \setcounter{page}{1}          

   \author{Chandra B. Singh
      \inst{1}
   \and Toru Okuda
      \inst{2}
   \and Ramiz Aktar
      \inst{3}
   }

   \institute{South-Western Institute for Astronomy Research, Yunnan University, University Town, Chenggong, Kunming 650500, People's Republic of China; {\it chandrasingh@ynu.edu.cn.}\\
        \and 
              Hakodate Campus, Hokkaido University of Education, Hachiman-Cho 1-2, Hakodate, Hokkaido 040-8567, Japan\\
        \and
              Department of Astronomy, Xiamen University, Xiamen, Fujian 361005, People's Republic of China\\}
           
\vs\no
   {\small Received~~20xx month day; accepted~~20xx~~month day}

\abstract{We study two dimensional low angular momentum flow  around the black hole using the resistive magnetohydrodynamic module 
of PLUTO code.  Simulations have been performed for the flows with parameters of specific angular momentum, specific energy, and 
magnetic field which may be expected for the flow around Sgr A*.
For flows with lower resistivity $\eta=10^{-6}$ and $0.01$, the luminosity and the shock location on the equator vary
quasi-periodically. The power density spectra of luminosity variation show the peak frequencies which correspond to the periods 
of 5 $\times 10^5$,  1.4 $\times 10^5$, and 5$\times 10^4$ seconds, respectively. 
These quasi-periodic oscillations (QPOs)  occur due to the interaction between the outer oscillating standing
shock and the inner weak shocks occurring at the innermost hot blob.
While for cases with higher resistivity $\eta=0.1$ and 1.0,  the high resistivity considerably suppresses the magnetic activity 
such as the MHD turbulence and the flows tend to be steady and symmetric to the equator.
The steady standing shock is formed more outward compared with the hydrodynamical flow.
The low angular momentum flow model with the above flow parameters and with low resistivity has a possibility for the explanation 
of the long-term flares with  $\sim$ one per day and $\sim$ 5 -- 10 days of Sgr A*  in the latest observations by Chandra, 
Swift, and XMM-Newton monitoring of Sgr A*.
\keywords{accretion, accretion disks --- magnetohydrodynamics (MHD) --- methods:numerical --- shock waves --- Galaxy: center}
}

   \authorrunning{Singh, Okuda \& Aktar}
   \titlerunning{Resistive advective flows with standing shocks}

   \maketitle

\section{Introduction}
Black hole accretion is the most efficient process which can address the issue of power generated in the neighbourhood of a black hole.
Historically, the study of black hole accretion has been based on two extreme cases of accretion process: radiatively inefficient 
flow called Bondi flow (\cite{bon52, mic72}) and radiatively efficient one called Keplerian disk (\cite{ss73, nt73}). Both suffer
from certain limitations.  Spherical Bondi flow with zero angular momentum is quite fast and  cannot explain the high luminosities
associated with observational signatures around the black hole. However, in reality, accretion flow is supposed to have some amount
of angular momentum associated with it. On the other hand, cold, thin, Keplerian disk cannot explain the issue of change of spectral
states and associated temporal variabilities and it is not applicable for the close region around the black hole as pressure gradient and
advective radial velocity terms are ignored.

Accretion flow onto the black hole is supposed to be supersonic at the event horizon and subsonic at a large distance as the accretion
flow approaches the speed of light at the horizon with sound speed being of lesser value. So the flow with angular momentum must pass
through at least one sonic point before plunging onto the black hole and should be advective (\cite{lt80}). In case of 
accretion onto a star, even a small angular momentum will stop the matter fall onto its surface because of the infinite potential barrier
associated with the Newtonian potential. Whereas in the case of a black hole, gravity always wins over centrifugal force because of higher-order terms 
(\cite{cha93}). Not only that, for given values of specific energy and specific angular momentum of accretion flow 
around the black hole, multiple sonic points may also exist with the possibility of standing shocks (\cite{fuk87, cha89}). There have been 
several interesting works which explored the solutions with standing shocks in pseudo-Newtonian  potential (\cite{pw80}) taking into
account various prescriptions for alpha parameter (\cite{cha96, bec08, kc13}). General relativistic solutions for inviscid 
(\cite{das15}) and viscous disks (\cite{ck16}) with standing shocks have also been reported. Even in the
presence of magnetic field, formation of standing shocks in accretion flows have been explored (\cite{tak06, fuk07}).

In recent time the model which has wide recognition is the advection dominated accretion flow (ADAF) (\cite{nar94, nar97}) solution which
takes care of inner boundary condition around the black hole however, has only one sonic point close to the black hole. It should be
noted that advective flow with multiple sonic points may not necessarily be ADAF-type especially when standing shock exists in the
accretion flow (\cite{cha96}). Overall, ADAF solutions occupy a small region of parameter space for given specific energy and specific angular momentum 
(\cite{lu99, kc13, kc14}). The need of a sub-Keplerian component was presented in addition to Keplerian one, the sub-Keplerian
component can undergo shock transition and form a hot, puffed up region like corona (\cite{ct95}). 
The properties of post-shock region formed from the natural course of flow dynamics can address issues like 
state transitions (\cite{mc10}), origin of hard power-tail and low frequency QPOs  
(\cite{cmd15}) and also the origin of outflows  (\cite{das01, sc11, akt15}). 

In the last 25 years, there have been a significant amount of simulation works dedicated to explore the formation of standing shock in
low angular momentum sub-Keplerian advective flows around black holes. Using smoothed particle hydrodynamics (SPH) simulations, 
stable standing shocks were shown to form in one-dimensional (\cite{cm93}) and two-dimensional setups (\cite{mlc94}) as predicted
by semi-analytical solutions of inviscid flows (\cite{cha89}). For the first time, the dependence of standing shock stability on 
values of viscosity parameters was shown by SPH simulations as well (\cite{cm95}). The origin of outflows from the post-shock region
in accretion disks were shown in simulations using SPH (\cite{mrc96}),  Eulerian total variation diminishing (TVD) (\cite{rcm97, otm07}) and Lagrangian 
TVD (\cite{lee16}). In the presence of cooling, the post-shock region may oscillate as the cooling time scale
becomes comparable to free-fall time scale and can be responsible for quasi-periodic oscillations(QPOs) in case of stellar mass
as well as supermassive black holes (\cite{msc96, otm07}). Besides the case of inviscid flow, viscosity can also
induce shock oscillations and give rise to QPOs (\cite{lan98, cam04, lan08, lee11, das14, lee16}). There have been some works regarding 
stability or instability of the shock and shocks seem to be stable against axisymmetric (\cite{nak92, nak94, noha94,
leke16}) as well as non-axisymmetric perturbations (\cite{mol99, gufo03, gulu06}). Recently it has established
through numerical simulations that advective flow can be segregated into two components, Keplerian as well as sub-Keplerian, 
in the presence of viscous heating and cooling processes (\cite{gc13, ggc15, rc17}). All the above-mentioned simulation works addressed 
the accretion flow behaviour around a non-rotating black hole using pseudo-Newtonian potential. Recently general
relativistic high-resolution shock-capturing simulation code was used to study the scenario in the Schwarzschild (\cite{kim17})
and Kerr (\cite{kim19}) space-time which further established the formation of standing shock in hydrodynamic flow around non-rotating 
as well as rotating black hole in full general relativistic treatment. 
However, till now there has been only one work taking into account different magnitudes of magnetic field strength in such flows in 
the presence of standing shocks (\cite{okuda19}). The long term evolution properties were investigated and long term flares in connection 
with Sgr A* could be explained.

Simulation works dealing with advective flows usually take into account two types of setup : torus equilibrium solution (e.g.,  
(\cite{sto01, mck02}) and Bondi flow along with arbitrary choice of specific angular momentum (\cite{pb03}). Our study
involves a third and different kind of set up where we take initial conditions based on exact solutions of hydrodynamic equations 
(\cite{cha89}). We are dealing with a big gap in parameter 
space which lie between the regime of high angular momentum torus and zero angular momentum Bondi flow. To make our study simpler, we 
are dealing with inviscid flows  having constant and small specific angular momentum value which are lower than that of Keplerian value of the specific angular momentum 
for an inner-most stable orbit. Such low angular momentum flows are likely to be present in binary systems accreting  winds from companion beside Roche lobe overflow as well as 
active galactic nuclei where winds from stellar clusters collide and lose angular momentum before getting accreted onto central black holes (\cite{ct95, sm01, sm02, mon06}). 
The objective of our work is to study the effects of resistivity with the varying magnitude on the formation and stability of standing shocks in low angular momentum 
accretion flows around the black holes which have not been explored before.

Section 2 shows details of 1.5 dimensional (1.5D) theoretical solutions which have been used for the simulation set up. In section 3, basic equations 
solved by simulation code are presented. Besides, the details of computational domain, initial, and boundary conditions are described 
in section 3. Section 4 contains details of numerical results followed by section 5 where we present a summary and discussion of our work.

\section{Theoretical solution}
\label {sec:theoretical solution}
We consider a semi-analytical approach of solving the standard conservation equations under hydrodynamic (HD) framework. The calculations 
are done in cylindrical co-ordinates with co-ordinates $R$ and $z$. Axi-symmetry is assumed for the angular $\phi$ co-ordinate. 
For simplicity, we further assume that the flow velocity along the vertical direction is zero and therefore only integrate along the radial 
co-ordinate assuming vertically averaged dynamical quantities. 

We define the scale radius as Schwarzschild radius $r_{g}$= $2GM/c^2$, with $M$ being the mass of the central compact object, G as the gravitational 
constant and $c$ is the speed of light. The matter that is accreted onto the central compact object has radial velocity given by $u_{\rm R}$, specific 
angular momentum $L$, and total specific energy $\cal E$. As the semi-analytical calculations are carried out using non-dimensional quantities, we define the following  -
\begin{equation}\label{eq:non-dim}
r = \frac{R}{r_g};  \,\,\,\,\, h =\frac{z}{r_g}; \,\,\,\,\, v_{\rm R} = \frac{u_{\rm R}}{c}; \,\,\,\,\, \lambda = \frac{L }{r_{g} c};  \,\,\,\,\,\, \epsilon = \frac{\cal E}{c^{2}}
\end{equation}
 
For studying ideal, inviscid flow onto a compact object, we deal with mass conservation equation,
\begin{equation}\label{eq:mass_acc}
  \dot M = 4 \pi \rho v_{\rm R} r h,
\end{equation}
where $\rho$ is the density and $h$ being the half-thickness of flow. 
Energy conservation gives us a relation of specific energy of the flow or Bernoulli constant,
\begin{equation}\label{eq:tot_E}
  \epsilon = \frac {v_{\rm R}^{2}} {2} + \frac {c_{s}^{2}} {\Gamma - 1} + \frac {\lambda^{2}}{2r^{2}} + \Phi.
\end{equation}
Here $c_{s}$ is adiabatic sound speed in units of $c$ and $\Gamma$ is the adiabatic index. 
$c_{s} = \sqrt \frac{\Gamma p}{\rho c^{2}}$.  $p$ is the thermal pressure and $\Phi$ is the non-dimensional 
gravitational potential given by  $-1/2(x-1)$ for non-rotating black hole (\cite{pw80}) where $x = r_{sp}/r_{g}$ and $r_{sp}$ being the spherical radius.  

Using vertical equilibrium condition, we evaluate the radial dependence for $h$
\begin{equation}\label{eq:ht_eq}
h = c_{s} \sqrt{x} (x - 1)
\end{equation}

For given values of $\epsilon$ and $\lambda$, we solve equations~(\ref{eq:mass_acc}) and (\ref{eq:tot_E}) and look for transonic conditions.
Differentiating the equations~(\ref{eq:mass_acc}) and (\ref{eq:tot_E}), we obtain

\begin{equation}\label{eq:cri_con}
  \frac{dv_{\rm R}}{dr}[v_{\rm R} - \frac{2c_{s}^2}{(\Gamma +1)v_{R}}] = \frac {2c_{s}^{2}}{\Gamma + 1} \frac{dlnf}{dr} - \frac{dG}{dr}, 
\end{equation}

where $G = \lambda^{2}/2r^{2} - 1/2(r-1)$ and $f = 2r^{3/2} (r-1)$ \\
(\cite{cha89}). 
At critical points, the vanishing of left-hand side gives radial velocity, 

\begin{equation}\label{eq:vel_cri}
(v_{\rm R})_{crit} = \sqrt \frac{2}{\Gamma + 1} (c_{s})_{crit},
\end{equation}

and the vanishing of right hand side gives sound speed, 

\begin{equation}\label{eq:sou_cri}
(c_{s})_{crit} = \frac{(\Gamma + 1)(r_{crit} - 2)}{r_{crit}^{2}} \frac{(\lambda_{K}^{2} - \lambda^{2})}{(5r_{crit} - 2)}.
\end{equation}

\begin{figure}
         \includegraphics[keepaspectratio, scale=0.4]{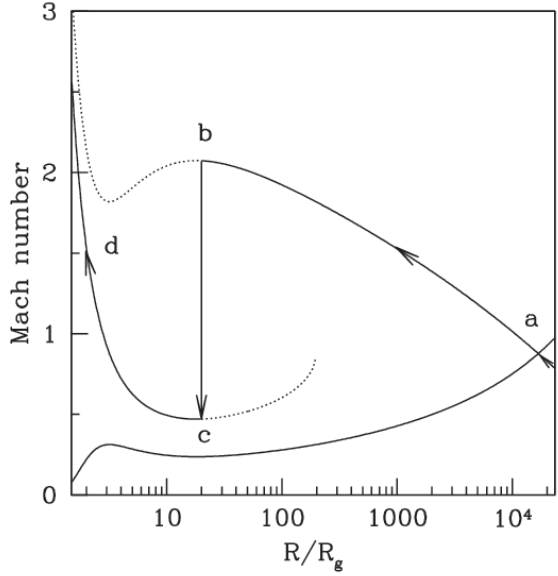}
\caption{Mach number versus radial distance from semi-analytical estimates for flow parameters, specific energy, $\epsilon$ = 1.98 $\times$ $10^{-6}$ and 
specific angular momentum, $\lambda = 1.35$, with $\Gamma$ = 1.6 (adopted from \cite{okuda19}).}
  \label{Fig: fig1}
  \end{figure}

The subscript $crit$ and $K$ represent quantities at the critical points and Keplerian orbits respectively.
In the case we obtain multiple critical points, we also check whether shock conditions are satisfied or not along the accretion flow. 
If shock conditions are satisfied, then there is a possibility of standing axisymmetric shock otherwise not. 
The shock location can be determined using an invariant quantity, $C$,  across the shock which is given by

\begin{equation}\label{eq:invar_sho}
C = \frac{[\mathcal{M_{+}}(3\Gamma - 1) +(2/\mathcal{M_{+}})]^{2}}{2 + (\Gamma - 1)\mathcal{M}_{+}^{2}} = 
\frac{[\mathcal{M_{-}}(3\Gamma - 1) +(2/\mathcal{M_{-}})]^{2}}{2 + (\Gamma - 1)\mathcal{M}_{-}^{2}} .
\end{equation}
Here, $\mathcal{M} = v_{\rm R}/c_{s}$ is the Mach number of the accretion flow. The subscripts $-$ and $+$ represent quantities in the pre-shock and post-shock region respectively.
Further details of the semi-analytical approach can be found in (\cite{cha89}). 

Fig.~\ref{Fig: fig1} shows the variation of Mach number, $\mathcal{M}$, of flow with radial distance from black hole obtained from exact 
theoretical solution solving conservation equations. The transonic flow passes through the outer critical point ``a" and continues its 
journey towards the black hole. The flow chooses to undergo shock transition along ``bc'', becomes subsonic then again accelerates 
towards the black hole and passes through inner critical point ``d'' to become supersonic before entering the black hole horizon. The shocked flow is 
preferrable in nature as the entropy generation is relatively higher compared to no shock flow.

\section{Numerical Setup}
\subsection{Basic Equations}
The numerical setup for the present work uses grid-based, finite volume computational fluid dynamics code, PLUTO 
(\cite{mig07, mig12}).  
Numerical simulations are carried out by solving the equations of classical resistive magnetohydrodynamics (MHD) in the conservative form:

\begin{eqnarray}
 \frac{\partial \rho} {\partial t} + \nabla \cdot \left(\rho \textbf{v}\right) &=& 0,\\
 \frac{\partial (\rho \textbf{v})}{\partial t} + \nabla \cdot [\rho \textbf{v} \textbf{v} - \textbf{B} \textbf{B}] + \nabla p_{t} &=& -\rho \nabla \Phi,\\
\frac{\partial E}{\partial t} + \nabla \cdot [(E + p_{t})\textbf{v}-(\textbf{v}\cdot \textbf{B}) \textbf{B} + \eta (\nabla \times \textbf{B}) \times \textbf{B}] &=& -\rho \textbf{v} \cdot \nabla \Phi, \\ 
\ \frac {\partial \textbf{B}}{\partial t} - \nabla \times (\textbf{v} \times \textbf{B} - \eta \nabla \times \textbf{B}) &=& 0. 
\end{eqnarray}

Here, $p_{t}$ being the total pressure with contribution from thermal pressure, $p$, and magnetic pressure, $B^{2}/2$. $E$ is the total energy density given by 
\begin{eqnarray}
E = \frac{p}{\Gamma-1} + \frac{1}{2}({\rho \textbf{v}^{2}} + \textbf{B}^{2}).
\end{eqnarray}
$\eta$ is the resistivity for which range of values have been chosen, $10^{-6}$, $0.01$, $0.1$ and $1$.
Vector potential $\textbf{A}$ is prescribed to generate the magnetic field $\textbf{B}$ as $\textbf{B} = \nabla \times \textbf{A}$. 
Following \cite{pb03}, the components of $\textbf{A}$ are as follows $A_{R} = 0$, $A_{\phi} = \frac{A_{0}z}{r_{sp}R}$ and $A_{z} = 0$.
Here, $A_{0} = sign(z) (\frac {8 \pi p_{out}}{\beta_{out}})^{1/2} R_{out}^{2}$ and $\beta_{out} = 8 \pi p_{out}/ B_{out}^{2}$, 
subscript ''out" denotes parameters at the outer boundary of the computational domain, $R_{out}$. Following \cite{okuda19}, we take a typical value of 5000 for $\beta_{out}$.

\begin{figure}
         \includegraphics[keepaspectratio, scale=0.4]{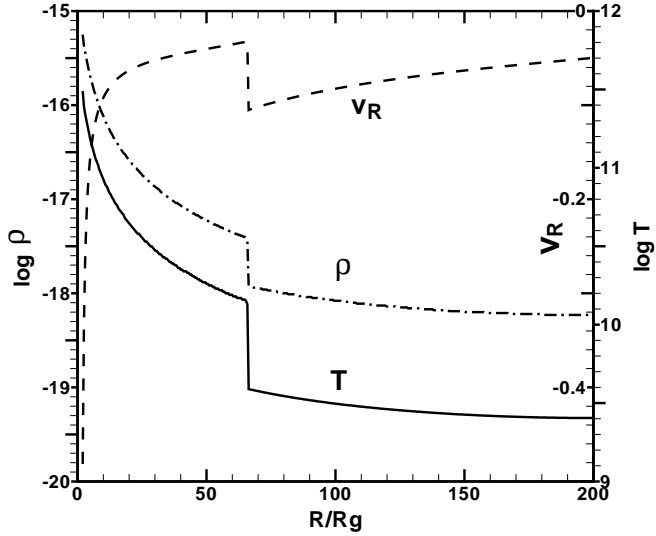}
\caption{Flow parameters on the equator, namely density ($\rho$ in g$cm^{-3}$), radial velocity ($v_{R}$) and temperature (T), 
for final state of HD simulation run. The standing shock is at $64.8 R_{g}$ (adapted from \cite{okuda19}). }
  \label{Fig: fig2}
  \end{figure}

\subsection{Initial and Boundary Conditions}
First, we have studied advective flows onto black holes in axisymmetric 2D cylindrical geometry ($R$, $z$) in 
HD framework. The theoretical solutions given in section~\ref{sec:theoretical solution} provide initial conditions of primitive variables, 
radial, and azimuthal velocity components, density, and pressure for 2D HD set up. 
Once the HD flow with the standing shock achieves a steady state, we use the solutions as the initial conditions for
 the magnetized flow with resistivity and let the simulation evolve further. 
The computational domain is $0 \leq R \le 200R_{g}$
and $-200R_{g} \le z \le 200R_{g}$ with the resolution of $1640 \times 820$ cells (for details, see \cite{okuda19}). 
Though we performed some simulation runs at half resolution, $820 \times 410$, and double resolution, $3280 \times 1640$, the results remain unchanged.\\
In both HD and MHD runs, the same boundary conditions are imposed. At the outer radial boundary, 
$R_{out}=200R_{g}$, there are two domains: the disk region where the matter is injected and the atmosphere
above the disk region. The flow parameters given by 1.5D theoretical solutions are provided in the region $-h_{out} 
\le z \le h_{out}$ where $h_{out}$ is the vertical equilibrium height at $R_{out}$. For the atmosphere region, the matter is 
allowed to leave the domain but not enter. The axisymmetric boundary condition is implemented at the inner radial boundary.
At $R=2R_{g}$, the absorbing condition is imposed in the computational domain. In the vertical direction, $z=\pm200R_{g}$, 
standard outflow boundary conditions are imposed. In the case of MHD run, the constant magnetic field is imposed on 
the outer radial boundary.
Fig.~\ref{Fig: fig2} shows profiles of density ($\rho$ in g$cm^{-3}$), radial velocity ($v_{R}$) and temperature(T) of the HD flow. 
The standing shock location from the simulation run is at $\sim 65 R_{g}$ which is 
significantly different from the predicted location from the theoretical solution, i.e. $\sim 20R_{g}$.
This is due to the assumption of the vertical hydrostatic equilibrium used in the 1.5D transonic solutions, which is
valid as far as the disk thickness $h$ is sufficiently small compared with the radius $r$ (that is, $h/r \ll$ 1).
However, in many cases of the low angular momentum flow with the standing shock, $h/r \sim $ 0.1 -- 0.5
because such flows are intrinsically advective and geometrically thick.
Therefore if the outer radial boundary is chosen to be very far from the predicted theoretical shock location,
 the difference between the numerical and theoretical shock locations becomes significant (\cite{okuda19}).

\section{Results and discussion}
To get the characteristic features of the flow,  we examine the time evolution of shock location $R_{s}$ on the
equator and total luminosity $L$ of the flow. 
The luminosity $L$ is calculated as follows, assuming that the gas is optically thin
\begin{eqnarray}
L = \int q_{ff} dV, 
\end{eqnarray}
where $q_{ff}$ represents the free-free emission rate per unit volume and integration is performed over all the whole
computational domain.
In order to have some estimate of how much matter is lost as an outflow from the system, mass outflow rate 
in the z-direction is given by,
\begin{eqnarray}
\dot M_{out} = \int_{0}^{R_{out}} 2\pi\rho(R, z_{out})v_{z}(R, z_{out})RdR - \int_{0}^{R_{out}} 2\pi\rho(R, -z_{out})v_{z}(R, -z_{out})RdR,
\end{eqnarray}
where $v_{z}$ is the vertical velocity.

\begin{figure*}
    \begin{tabular}{cc}

      \begin{minipage}{0.5\linewidth}
        \centering
        \includegraphics[keepaspectratio, scale=0.3]{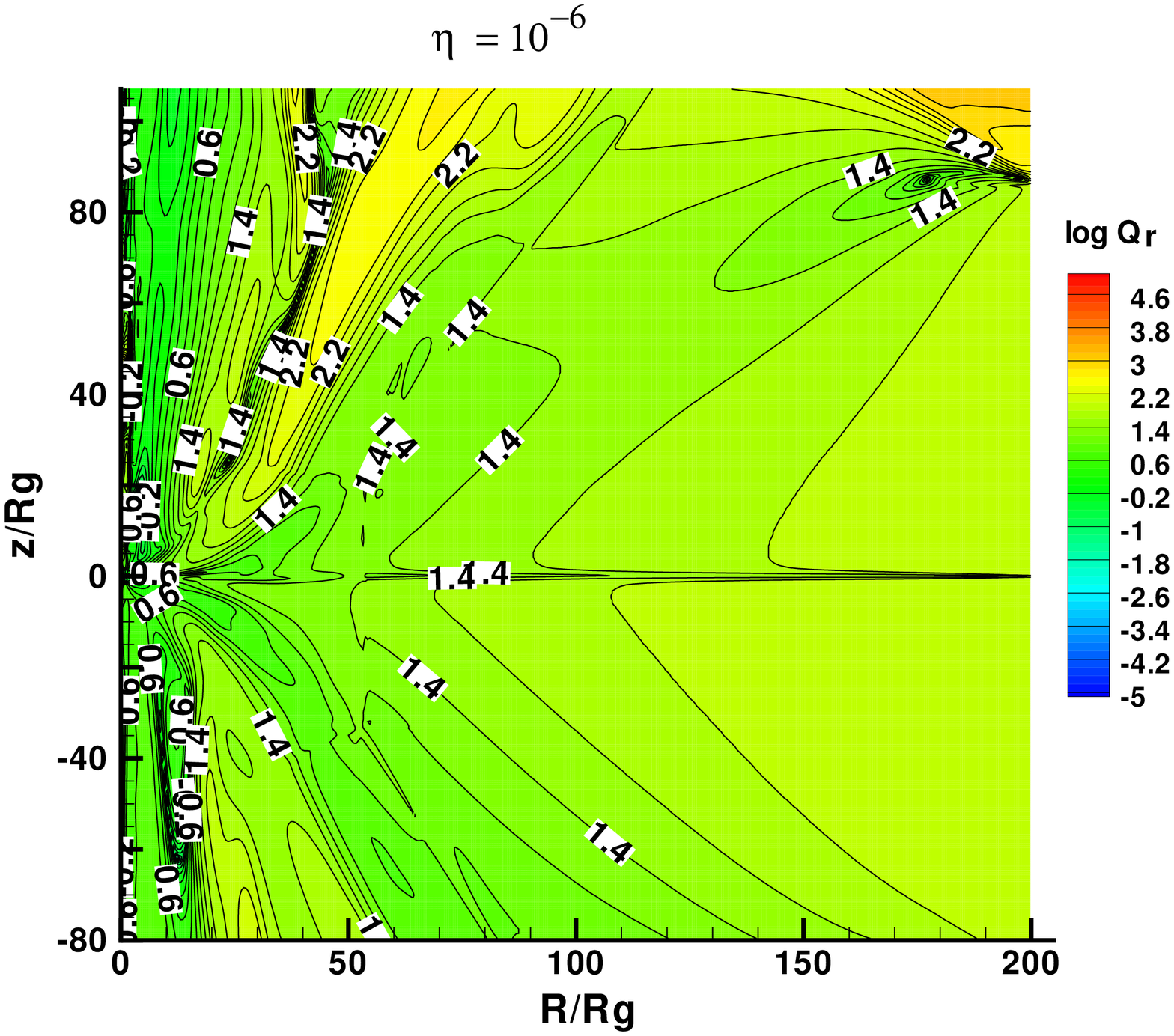}
        \label{fig:3a}
      \end{minipage}

      \begin{minipage}{0.5\linewidth}
        \centering
        \includegraphics[keepaspectratio, scale=0.3]{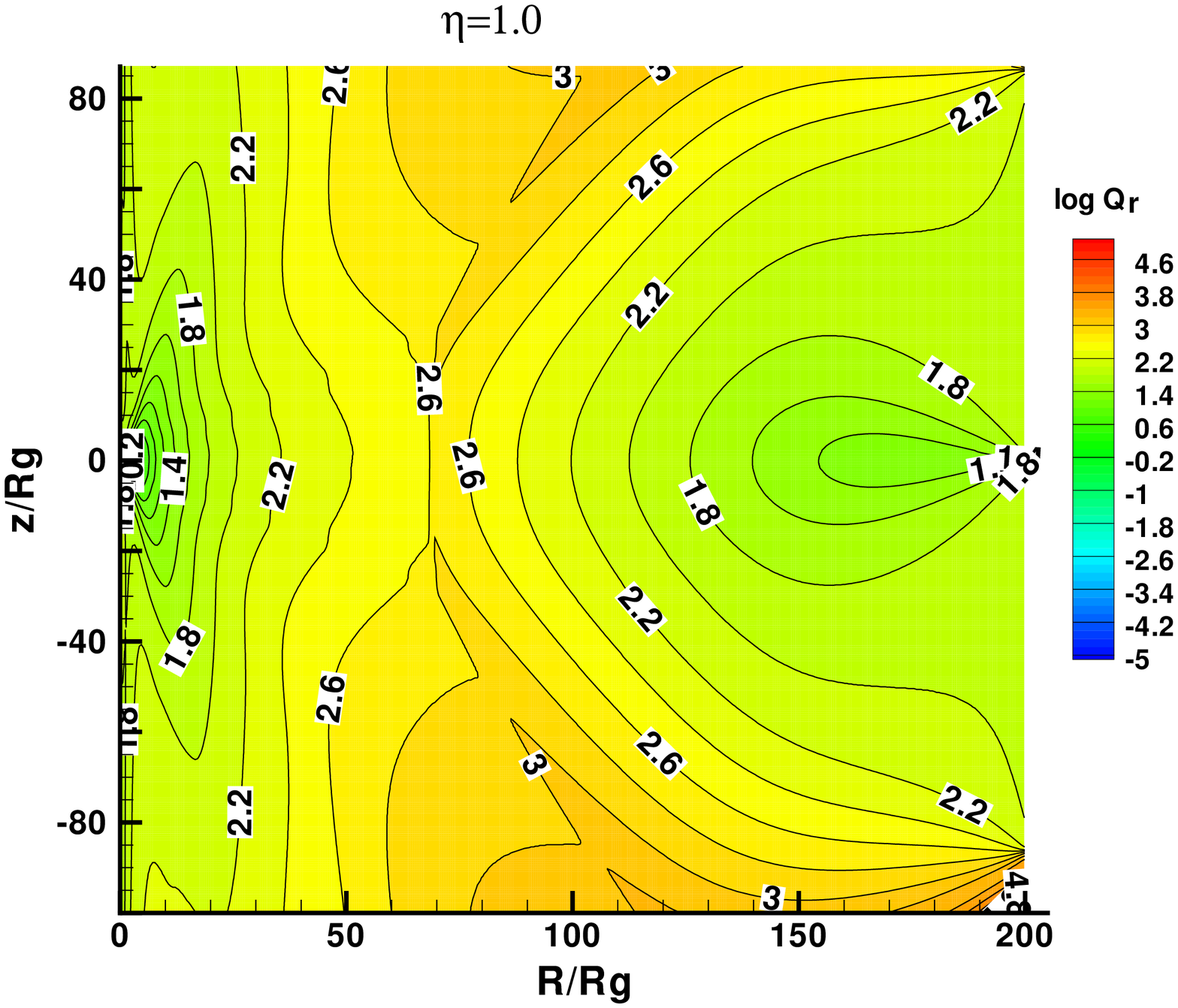}
        \label{fig:3b}
      \end{minipage} \\

    \end{tabular}
\caption {2D contours of  MRI-criterion $Q_{r}$ for cases of $\eta = 10^{-6}$ and 1.0 at times $t$ = $7 \times 10^{6}$ and $8.7 \times 10^{6}$ seconds, respectively.  For both cases,  $Q_{r} \gg $ 1 in most regions. The contours are asymmetric to the
 equator in the former but symmetric in the latter. 
  }
  \label{Fig: fig3}
  \end{figure*}

 In the present study, the time variability of the flow will correlate with the magnetorotational instability (MRI).
  We check whether the flow is subject to the MRI and whether we can resolve the fastest growing MRI mode or not.
 The stringent diagnostics of spatial resolution for the MRI instability has been examined in 
  3D magnetized flow.  Therefore, its application to our 2D magnetized flow may be limited.
 The critical wavelength of the instability mode is given by $\lambda_{\rm c} = 2\pi v_{\rm A}/\sqrt{3}\Omega$,
 where $v_{\rm A}$ and $\Omega$ are the Alfven velocity and the angular velocity (\cite{hawley91,balbus98}).
 A criterion value $Q_{\rm x}$ of the MRI resolution is defined by
 \begin{equation}
    Q_{\rm x}= \frac{\lambda_{\rm c}} {\Delta x},
  \end{equation}
 where $\Delta x$ is the mesh sizes $\Delta R$ and $\Delta z$ in the radial and vertical directions, respectively.
 When  $Q_{\rm x} \gg $ 1, the flow is unstable against the MRI instability, otherwise, the flow is stable.
Fig.~\ref{Fig: fig3} shows 2D contours of  radial MRI-criterion $Q_{r}$ for cases of $\eta = 10^{-6}$ and 1.0 at times 
$t$= $7 \times 10^{6}$ and $8.7 \times 10^{6}$ seconds, respectively. The analyses show  $Q_{r} \gg $ 1 in most regions for both cases.
So, both flows are unstable to the MRI. The contours are asymmetric to the equator  in the former but symmetric in the latter.

\subsection{Luminosity, shock location and mass in/outflow rates}

\begin{figure*}
    \begin{tabular}{cc}
      \begin{minipage}{0.5\linewidth}
        \centering
        \includegraphics[keepaspectratio, scale=0.45]{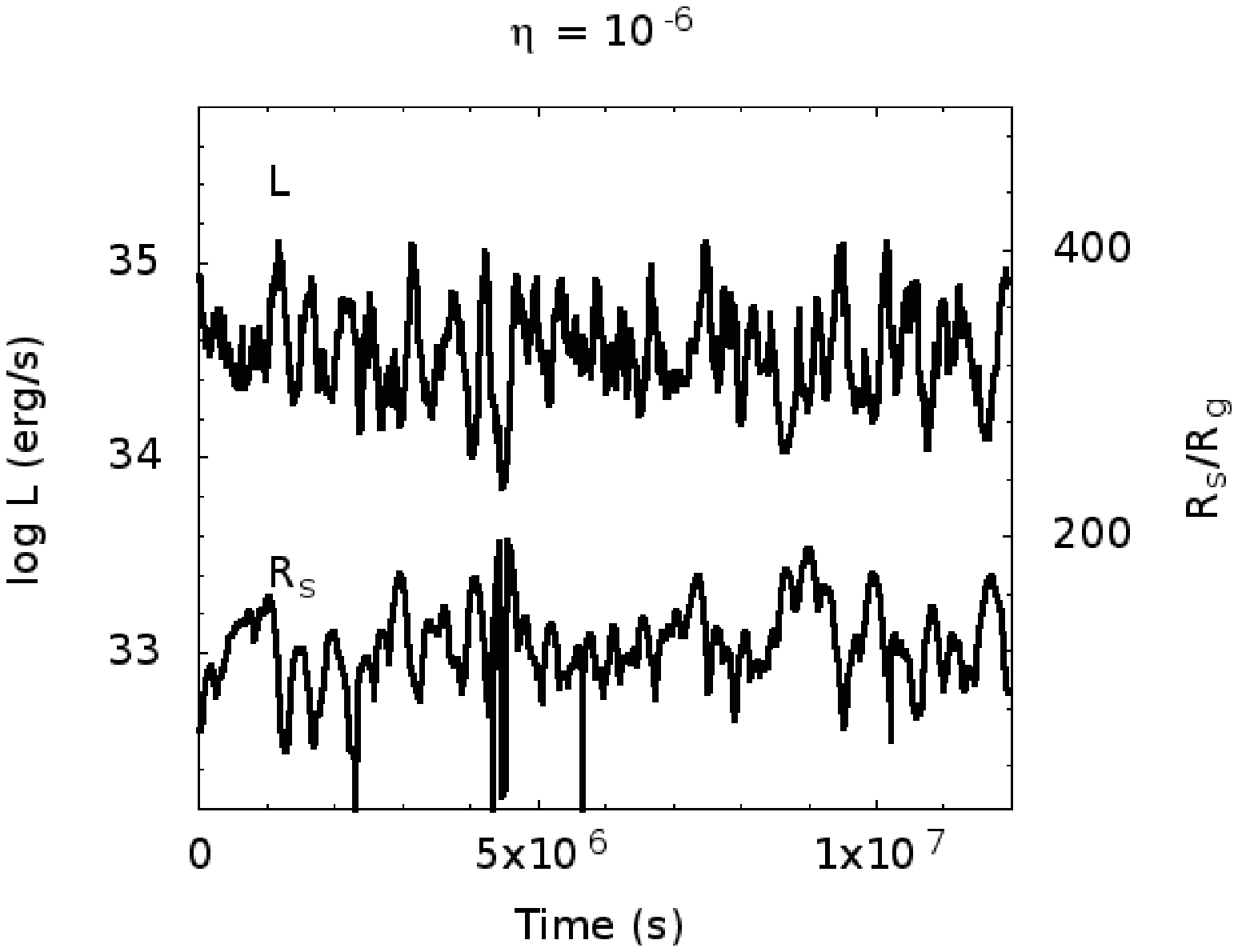}
        \label{fig:4a}
      \end{minipage}

      \begin{minipage}{0.5\linewidth}
        \centering
        \includegraphics[keepaspectratio, scale=0.45]{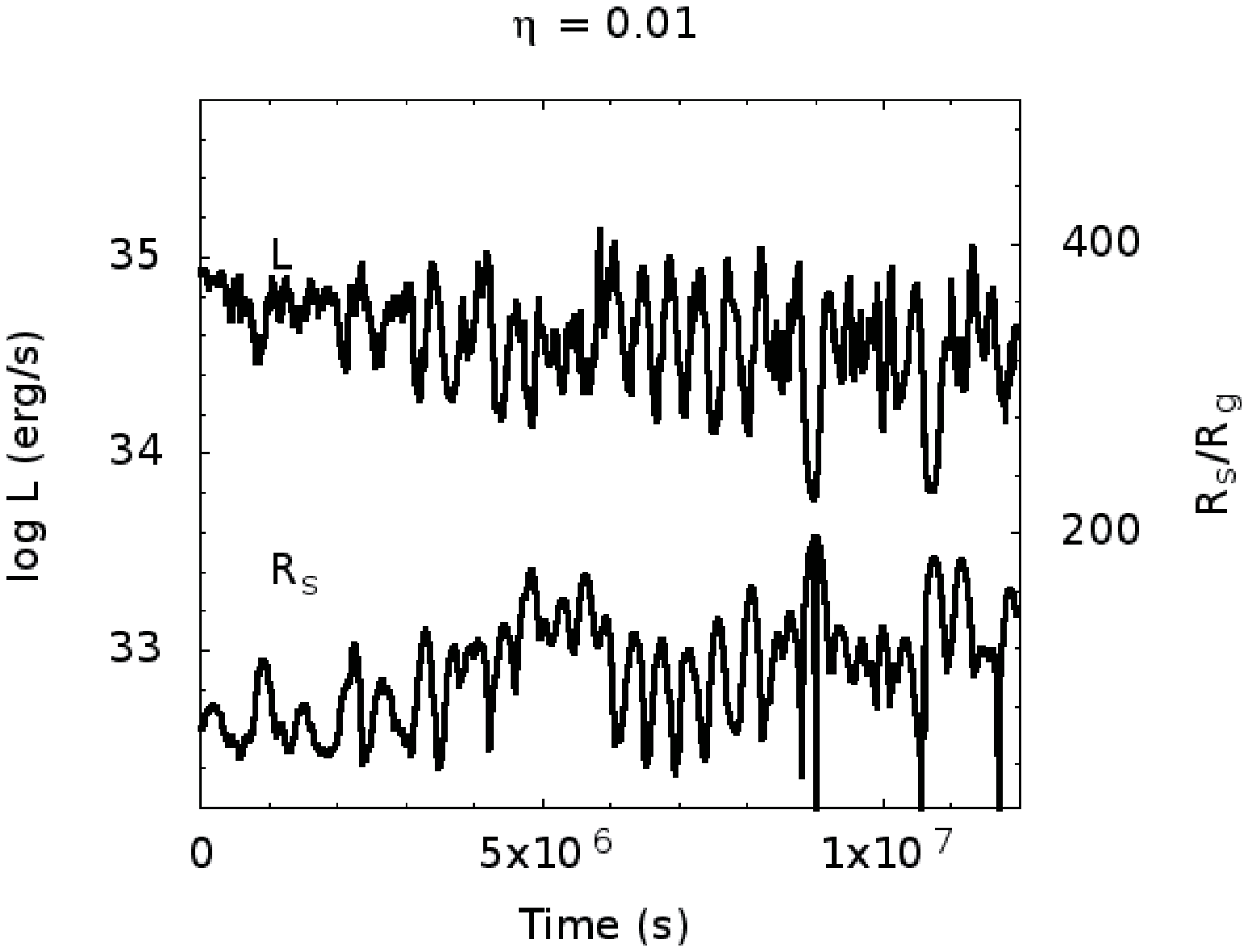}
        \label{fig:4b}
      \end{minipage} \\

  \begin{minipage}{0.5\linewidth}
        \centering
        \includegraphics[keepaspectratio, scale=0.45]{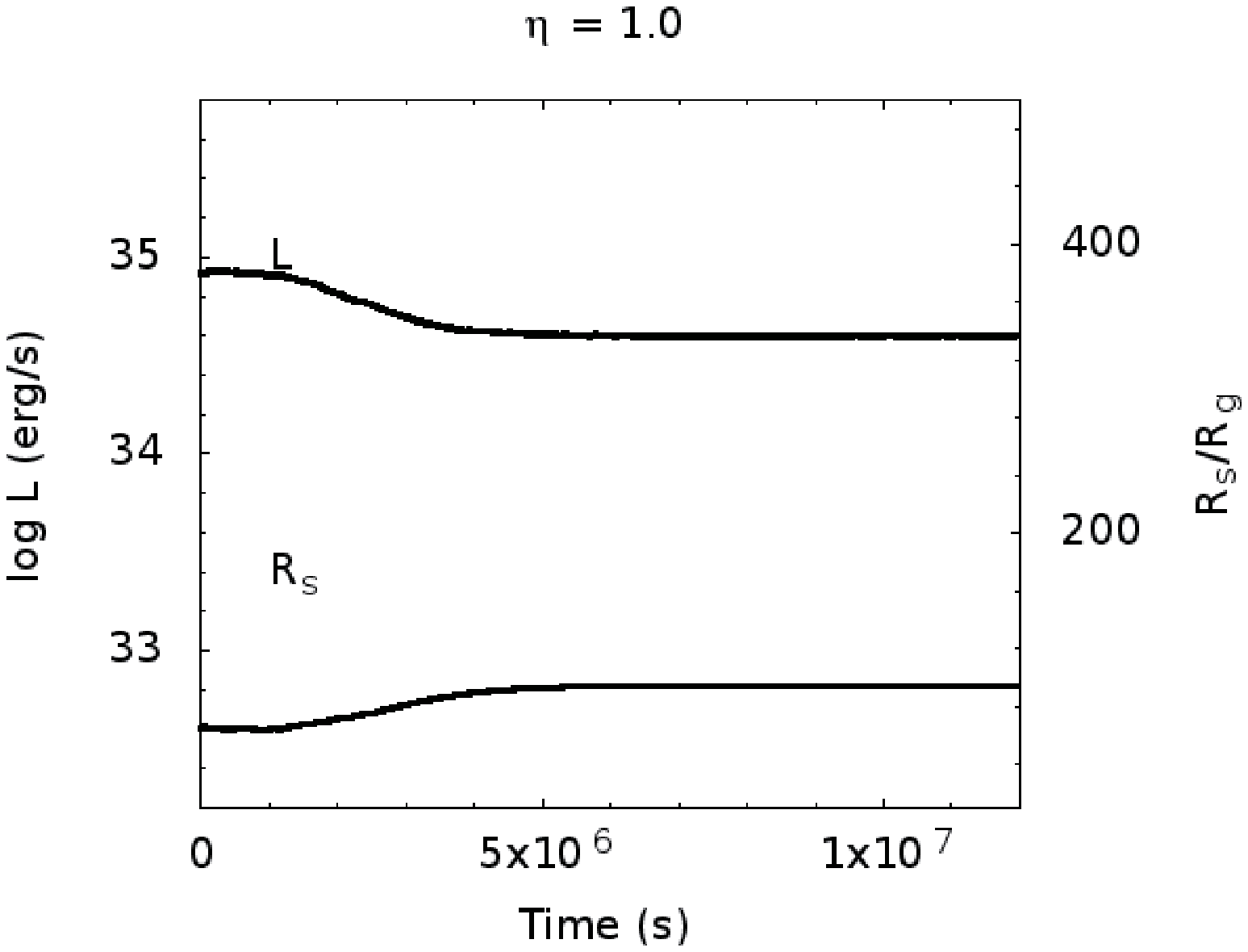}
        \label{fig:4d}
      \end{minipage}

  \begin{minipage}{0.5\linewidth}
        \centering
        \includegraphics[keepaspectratio, scale=0.45]{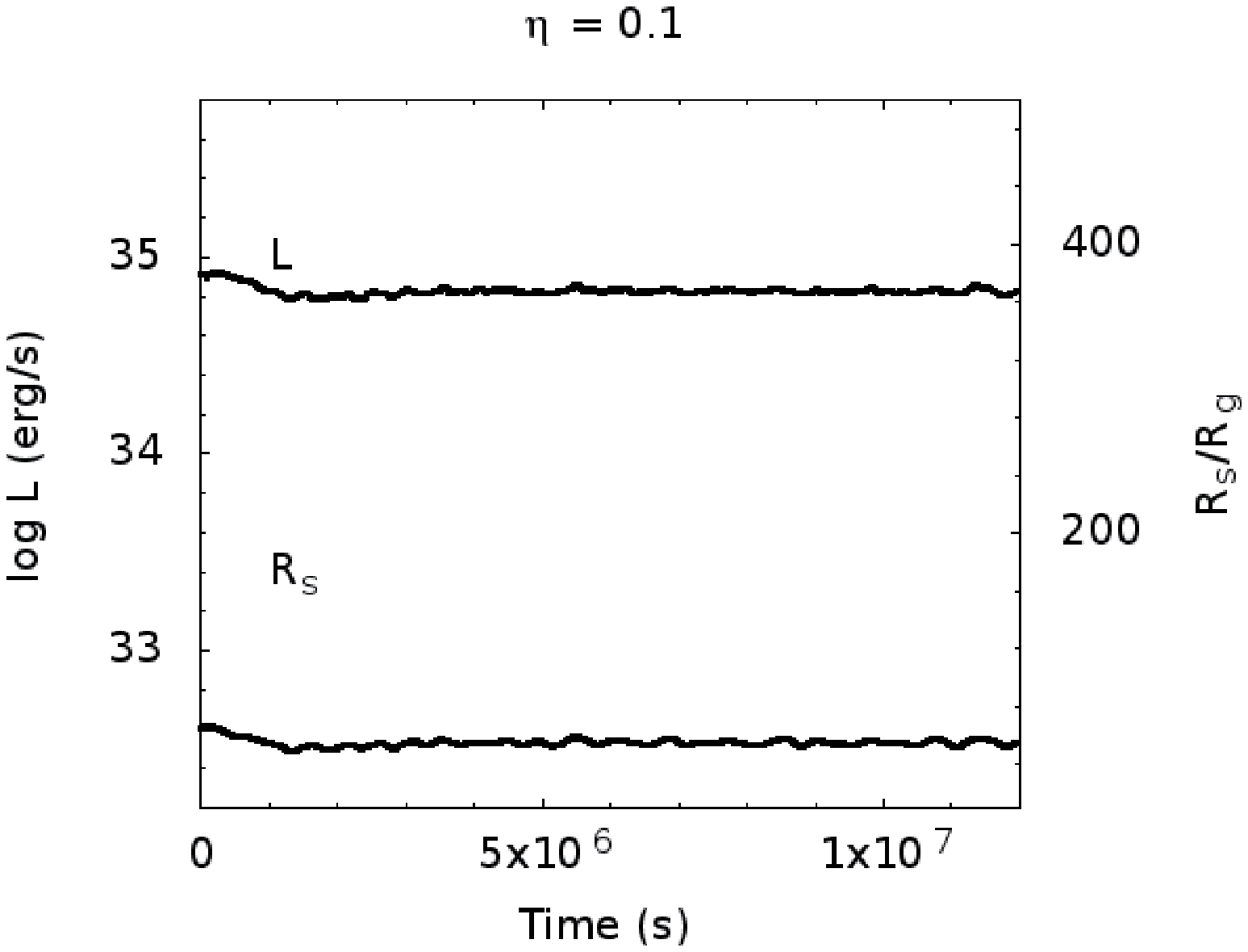}
        \label{fig:4c}
      \end{minipage}
    \end{tabular}
 \caption{Variation of Luminosity (L) and shock location ($R_{s}$) with time for resistive MHD flow with different values of resistivity,
$\eta$ = $10^{-6}$, 0.01, 0.1 and 1 (in clockwise direction).}
 \label{Fig: fig4}
  \end{figure*}

Fig.~\ref{Fig: fig4} shows how luminosity $L$ and shock location $R_{s}$ vary with time for different levels of resistivity in the flow. 
For lower values of resistivity, $\eta = 10^{-6}$ and $0.01$, there are features of irregular oscillation in the luminosity $L$ and the 
standing shock location $R_{s}$. As the shock moves towards the black hole, there is an increase in luminosity while the luminosity decreases when the shock recedes away. 
The shock and the luminosity oscillate irregularly with  time scales of $\sim 10^5 - 10^6$ s, 
and the luminosity varies maximumly by a factor of ten around the average $L$ is $\sim 3.0 \times 10^{34}$ erg s$^{-1}$.
On the other hand, for relatively high resistivity, $0.1$ and $1.0$, the oscillatory nature disappears and $L$ and $R_{s}$
 show small modulations or almost steady value at later times. The highly resistive flows behave qualitatively similar to that of HD flow.

\begin{figure*}
    \begin{tabular}{cc}

      \begin{minipage}{0.5\linewidth}
        \centering
        \includegraphics[keepaspectratio, scale=0.45]{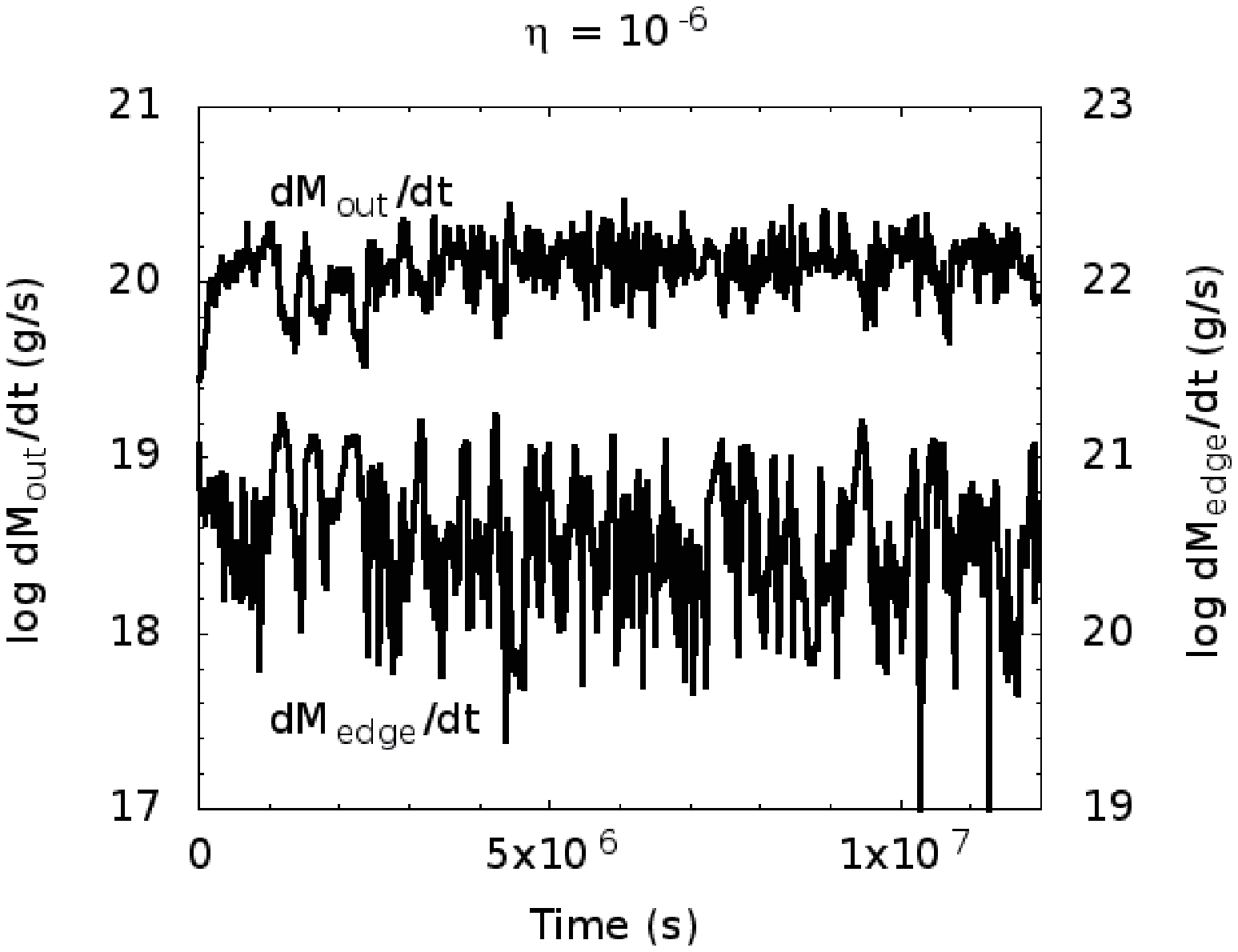}
        \label{fig:5a}
      \end{minipage}

      \begin{minipage}{0.5\linewidth}
        \centering
        \includegraphics[keepaspectratio, scale=0.45]{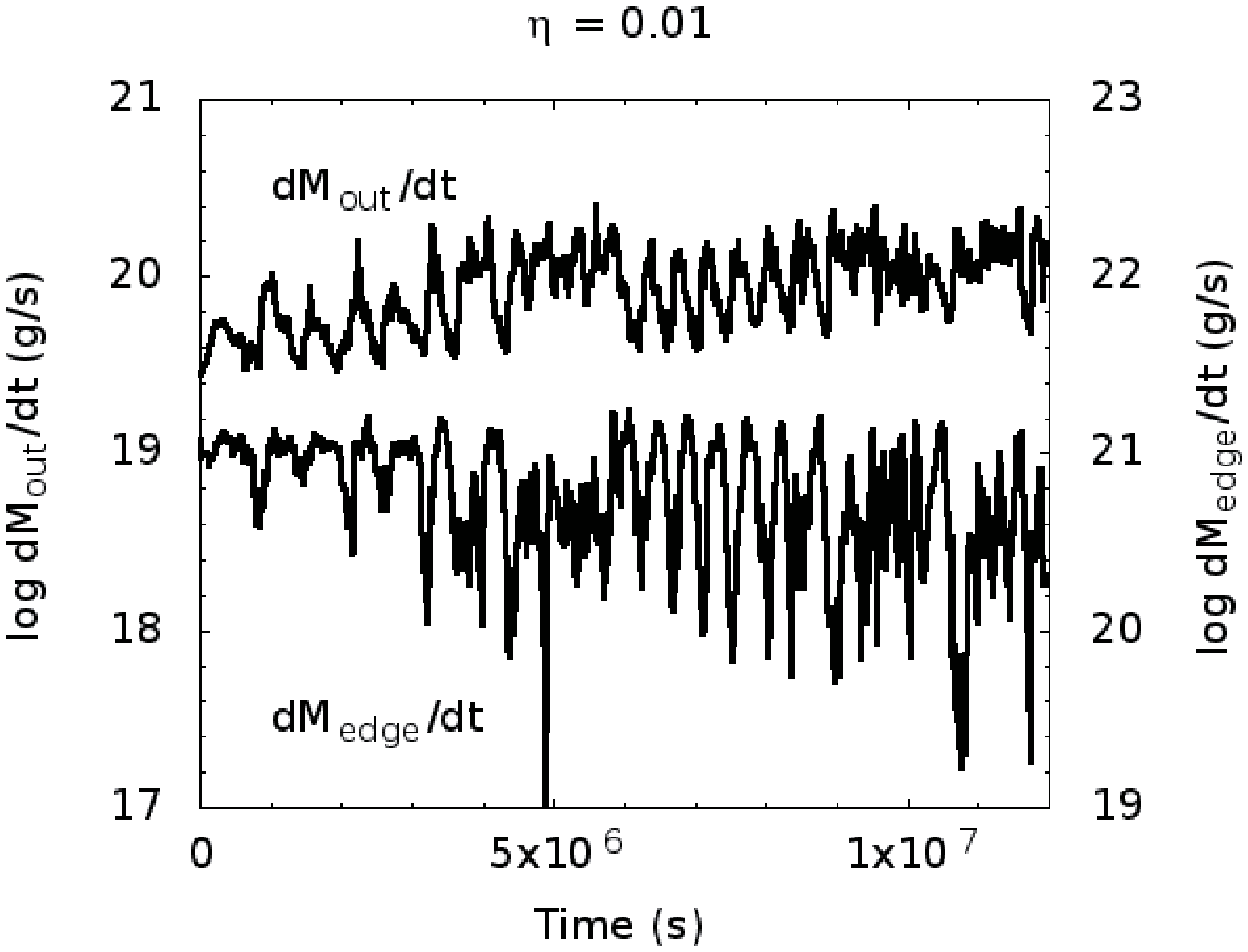}
        \label{fig:5b}
      \end{minipage} \\

  \begin{minipage}{0.5\linewidth}
        \centering
        \includegraphics[keepaspectratio, scale=0.45]{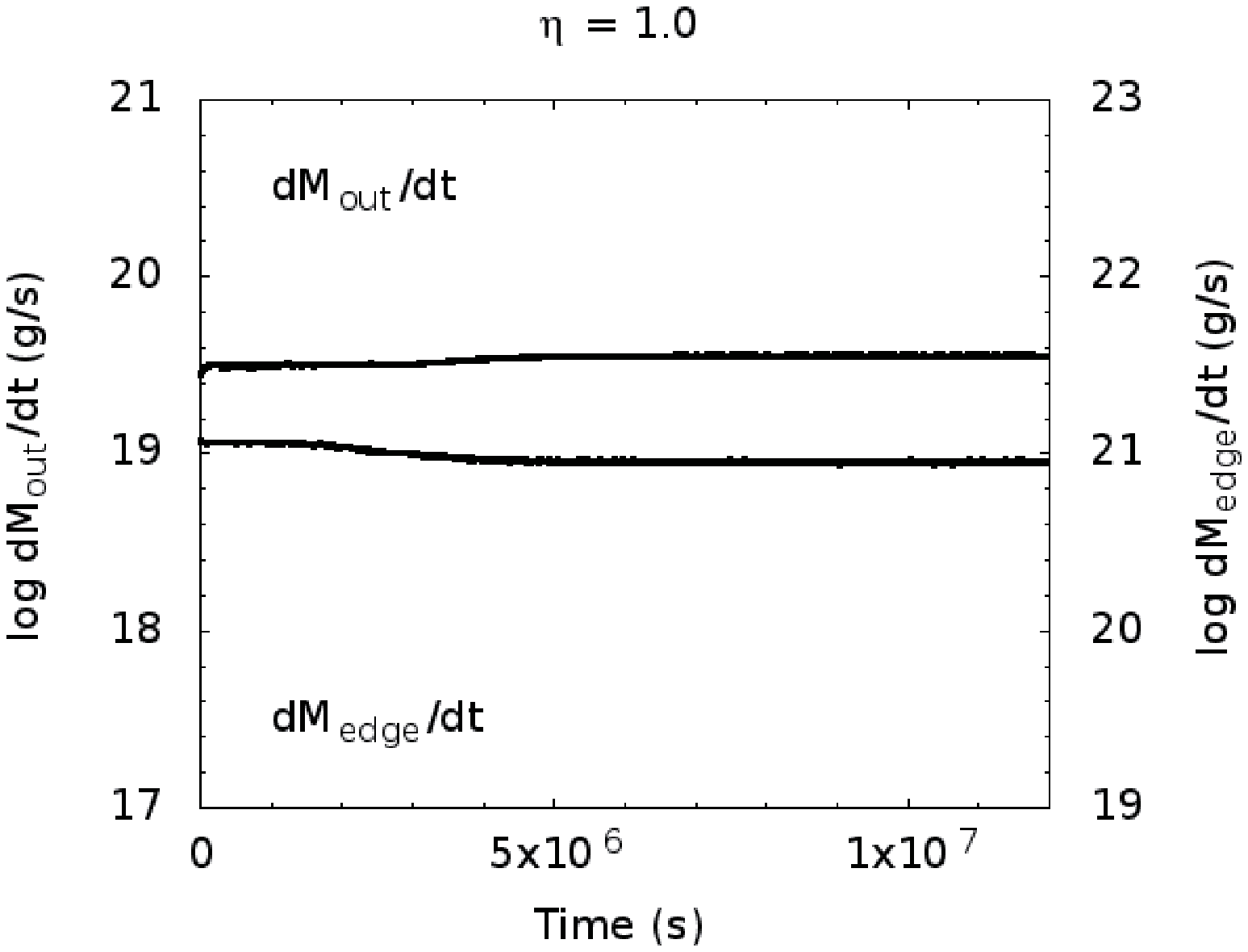}
        \label{fig:5d}
      \end{minipage}

  \begin{minipage}{0.5\linewidth}
        \centering
        \includegraphics[keepaspectratio, scale=0.45]{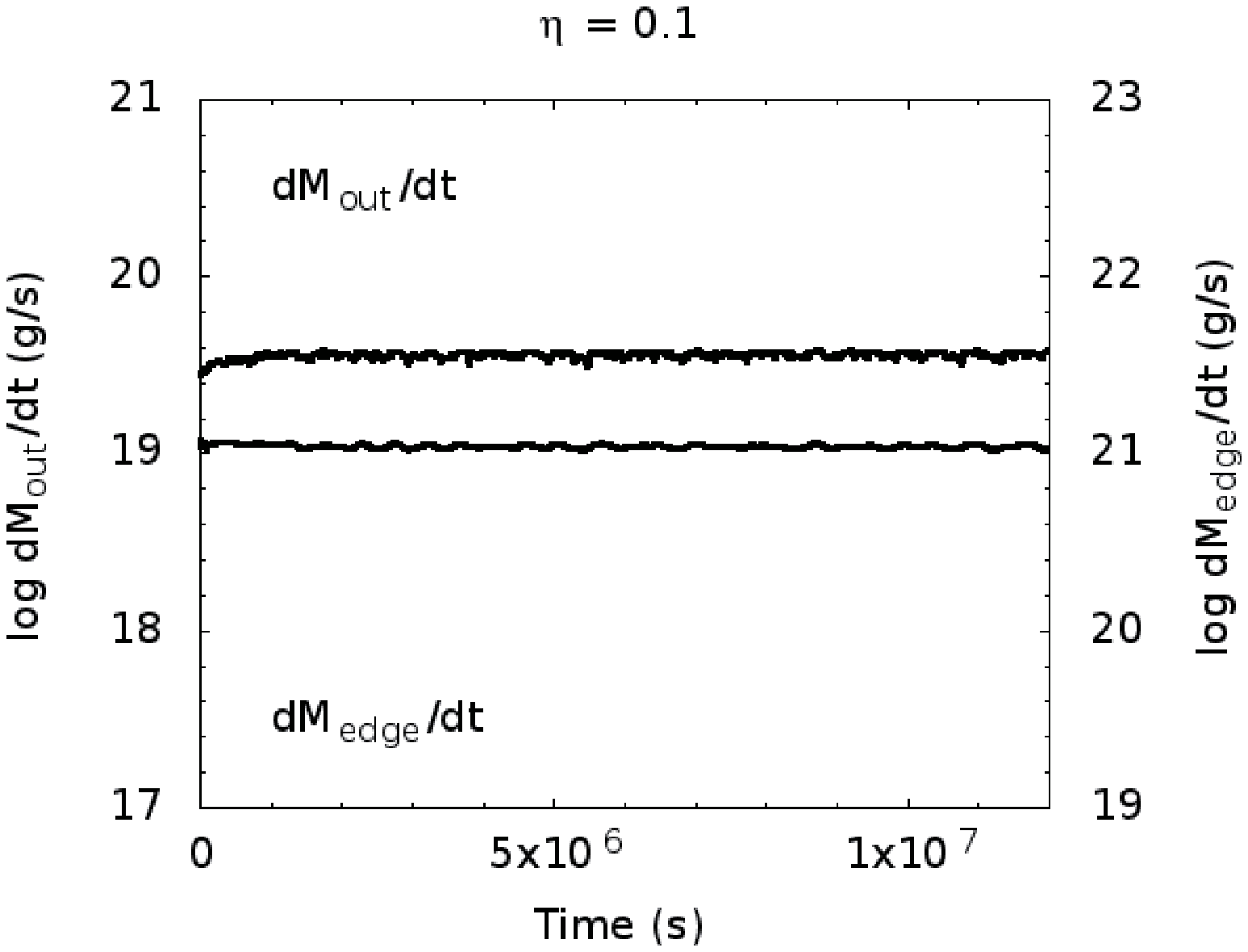}
        \label{fig:5c}
      \end{minipage}
    \end{tabular}
 \caption{Mass inflow ($\dot M_{edge}$) as well as outflow ($\dot M_{out}$) rate evolving with time for resistive MHD flow with different values of resistivity,
 $\eta$ = $10^{-6}$, 0.01, 0.1 and 1 (in clockwise direction).}
 \label{Fig: fig5}
  \end{figure*}

The mass inflow $\dot M_{edge}$ and outflow $\dot M_{out}$ rates are presented in Fig.~\ref{Fig: fig5} correspondings to 
 Fig.~\ref{Fig: fig4}. 
Similar to our previous work \cite{okuda19}, it has been established that there is a correlation between $L$ and $\dot M_{edge}$ and between 
$R_{s}$ and $\dot M_{out}$. While the variation of $L$ seems to be opposite in behaviour compared to $R_{s}$. 
That means when the post-shock region shrinks, the emission increases, and vice versa.
Since the low angular momentum flows are very advective, most of the input gas  $\dot M_{input}$
 ($ \sim 3 \times 10 ^{20}$ g s$^{-1}$) falls into the event horizon and $\dot M_{edge}$ is comparable to $\dot M_{input}$
 in all cases. However, the mass outflow rate $\dot M_{out}$ in low resistivity case is considerably high as a few tens of
 percent of the input accretion rate but in the high resistive case with $\eta$ = 0.1 and 1.0, $\dot M_{out}$ amounts
to $\sim$ 10 percent.  Such mass outflow rates are very high compared with the mass outflow rate found
in the usual accretion flow. It should be noted that the very high mass outflow rate in the low resistivity case
 may be correlated to the MRI turbulence, that is, the MRI turbulence plays important roles not only in the outward
 transfer of the angular momentum but also in outward mass transfer.

\begin{figure*}
    \begin{tabular}{cc}

      \begin{minipage}{0.5\linewidth}
        \centering
        \includegraphics[width=60mm,height=40mm]{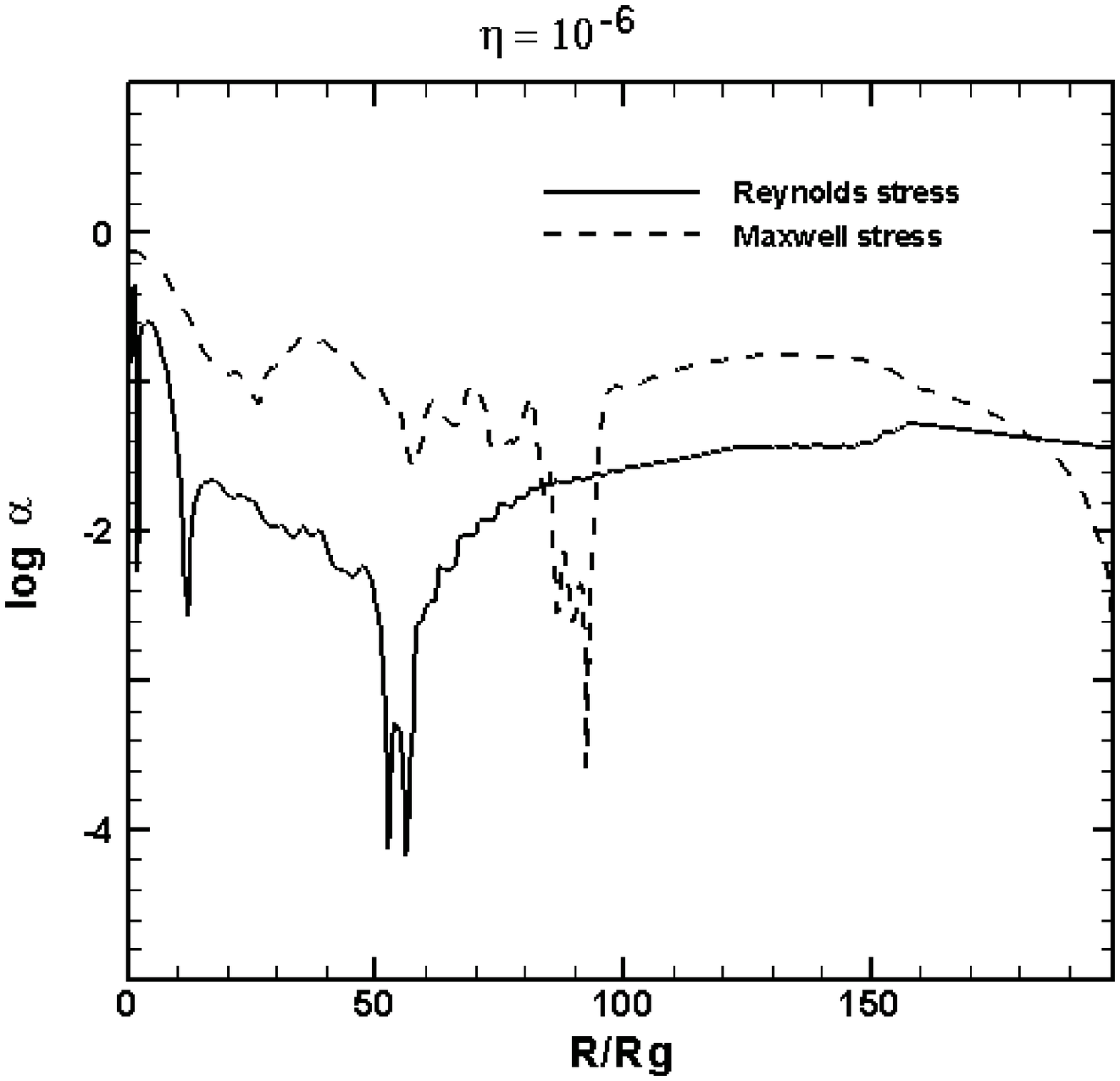}
        \label{fig:6a}
      \end{minipage}

      \begin{minipage}{0.5\linewidth}
        \centering
        \includegraphics[width=60mm,height=40mm]{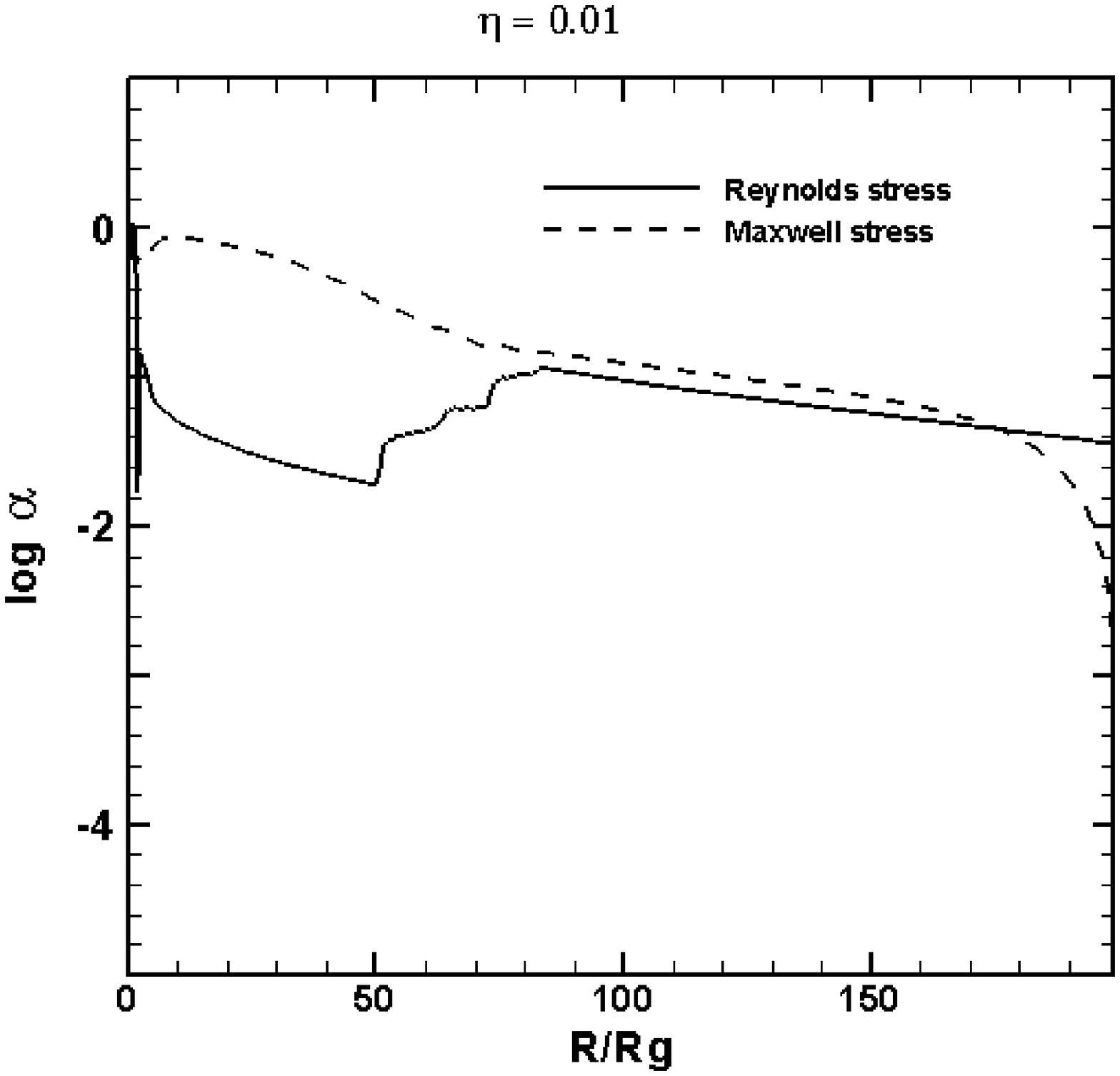}
        \label{fig:6b}
      \end{minipage} \\

  \begin{minipage}{0.5\linewidth}
        \centering
        \includegraphics[width=60mm,height=40mm]{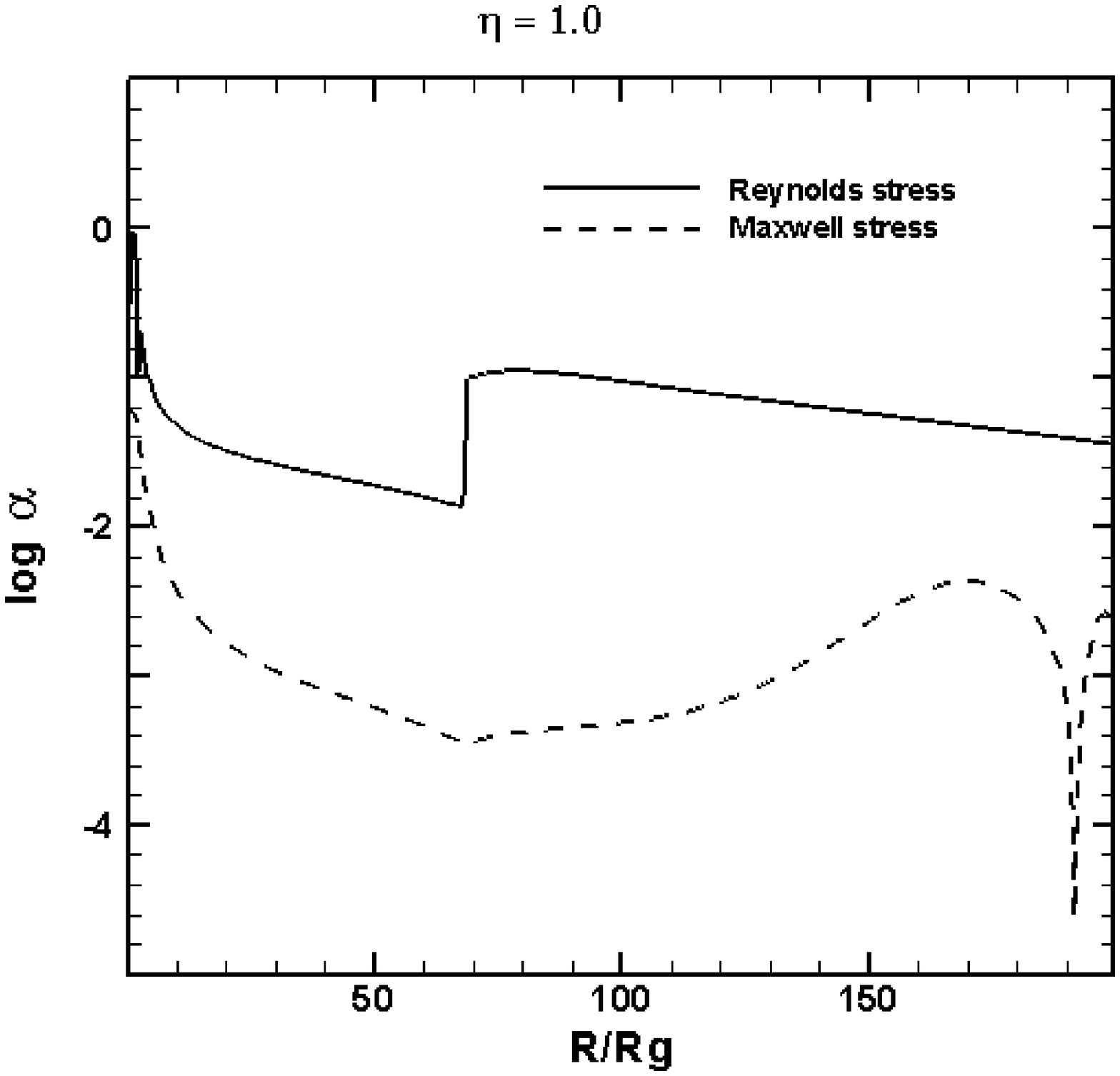}
        \label{fig:6d}
      \end{minipage}

  \begin{minipage}{0.5\linewidth}
        \centering
        \includegraphics[width=60mm,height=40mm]{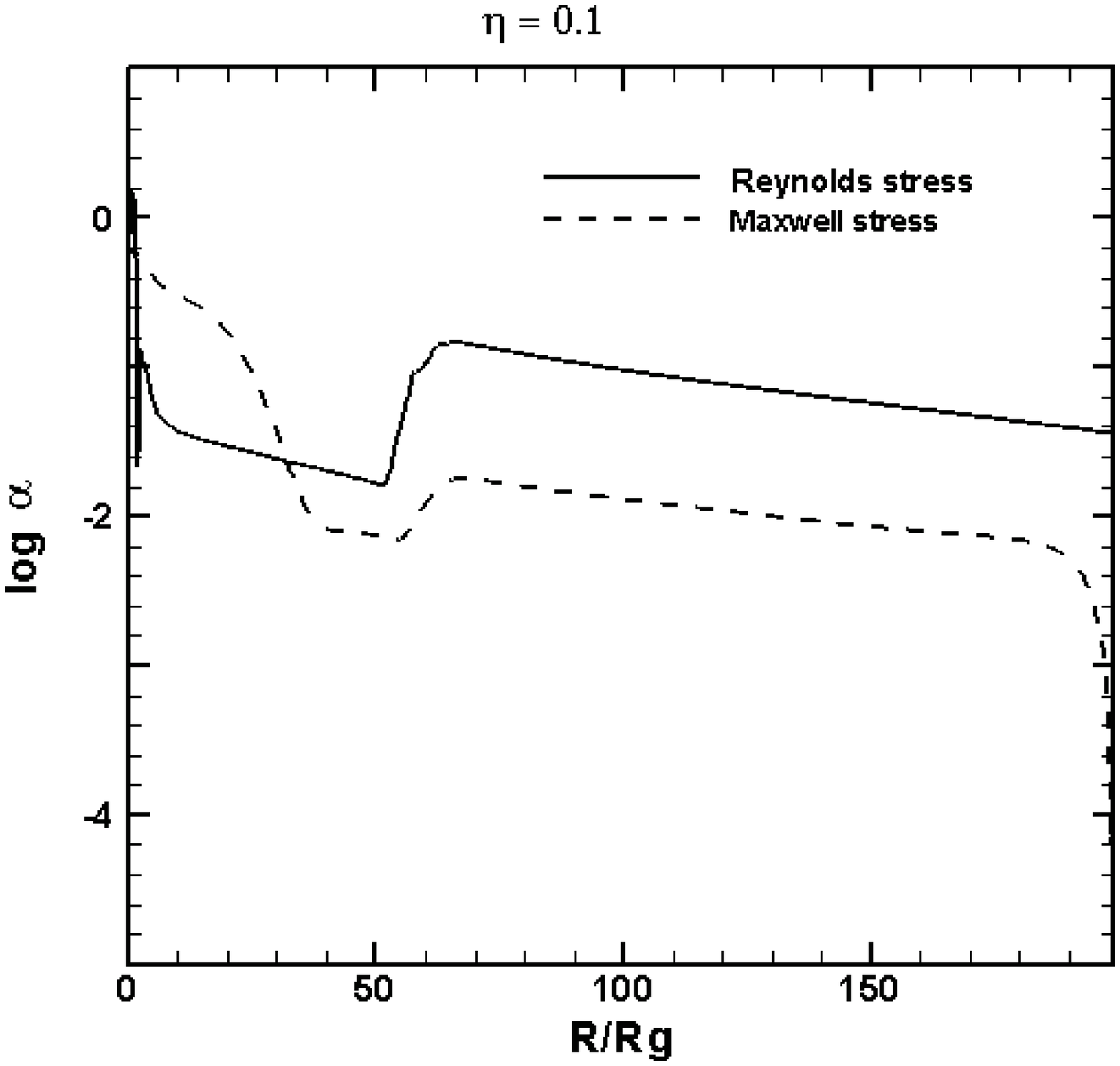}
        \label{fig:6c}
      \end{minipage}
    \end{tabular}
\caption{Radial profiles of normalized Reynolds stress $\alpha_{gas}$ and normalized Maxwell stress $\alpha_{mag}$  for resistive MHD flow with 
different resistivity, $\eta$ = $10^{-6}$, 0.01, 0.1 and 1 (in clockwise direction).
These values are space-averaged (between $-2R_{g}$ and $2R_{g}$ in z-direction) and time averaged (between $1.1 \times 10^{7}$ and $1.2 \times 10^{7}$ seconds).
}
  \label{Fig: fig6}
  \end{figure*}

\begin{figure*}
    \begin{tabular}{cc}

      \begin{minipage}{0.5\linewidth}
        \centering
        \includegraphics[width=60mm,height=50mm]{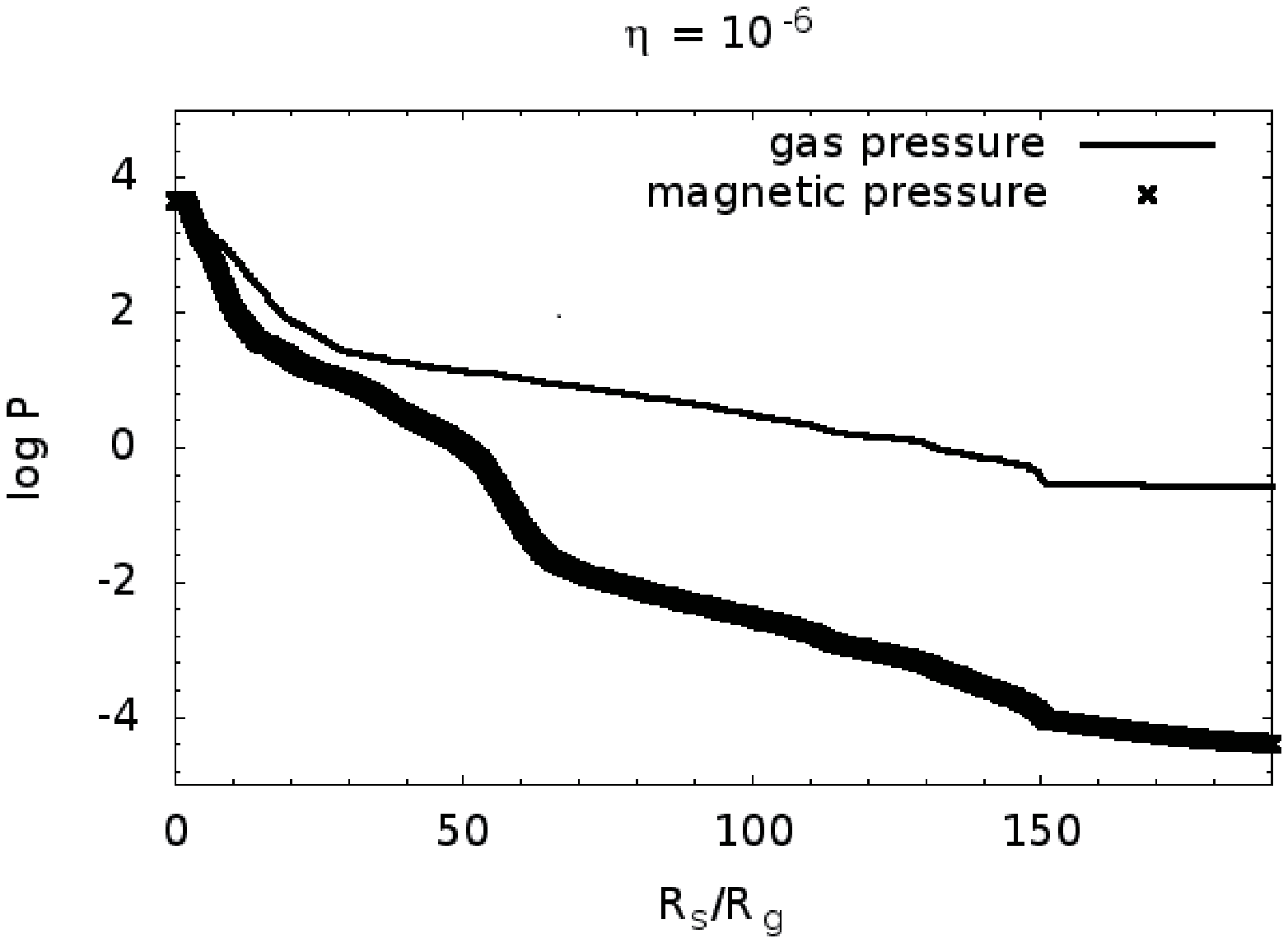}
        \label{fig:7a}
      \end{minipage}

      \begin{minipage}{0.5\linewidth}
        \centering
        \includegraphics[width=60mm,height=50mm]{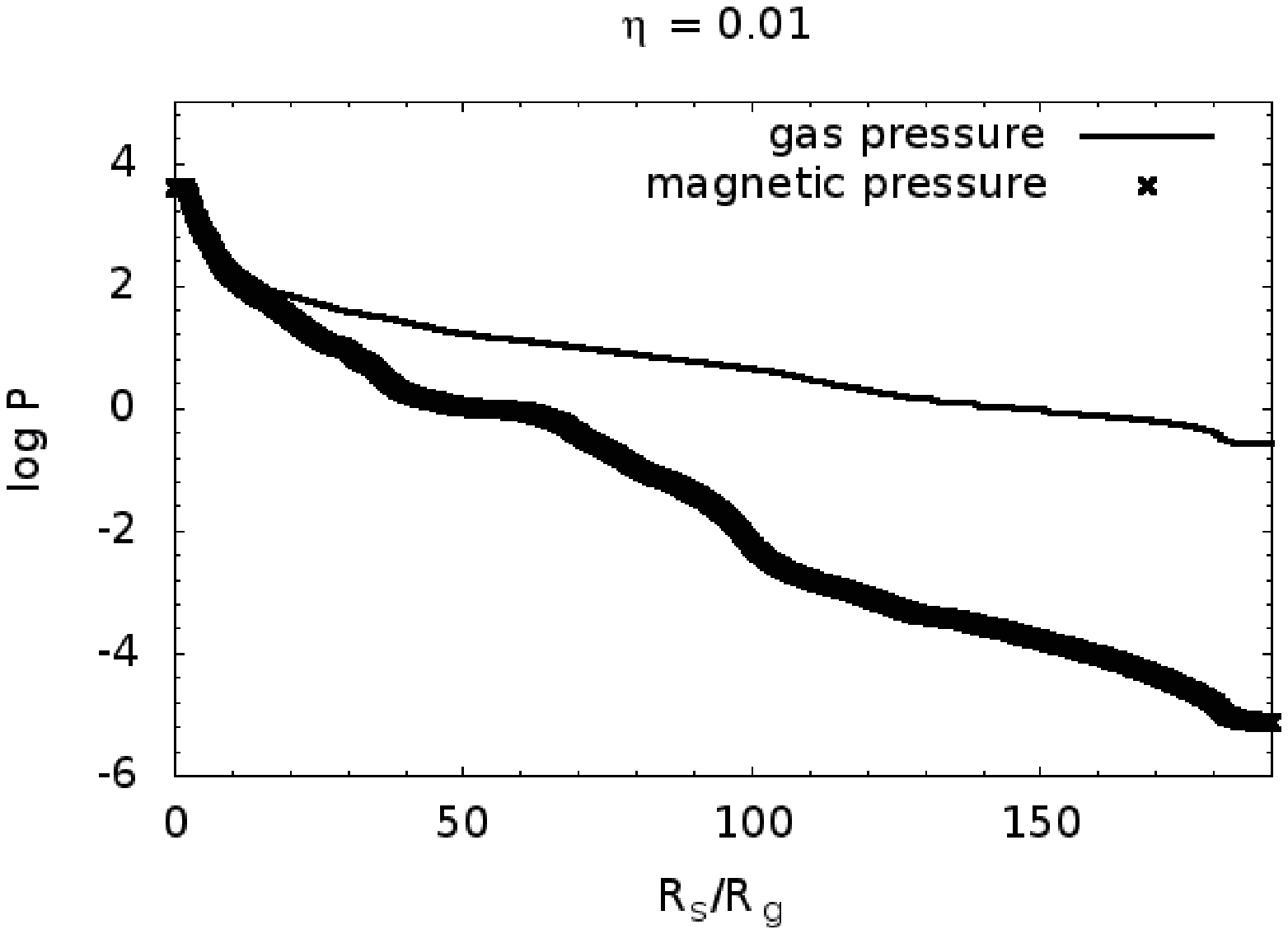}
        \label{fig:7b}
      \end{minipage} \\

  \begin{minipage}{0.5\linewidth}
        \centering
        \includegraphics[width=60mm,height=50mm]{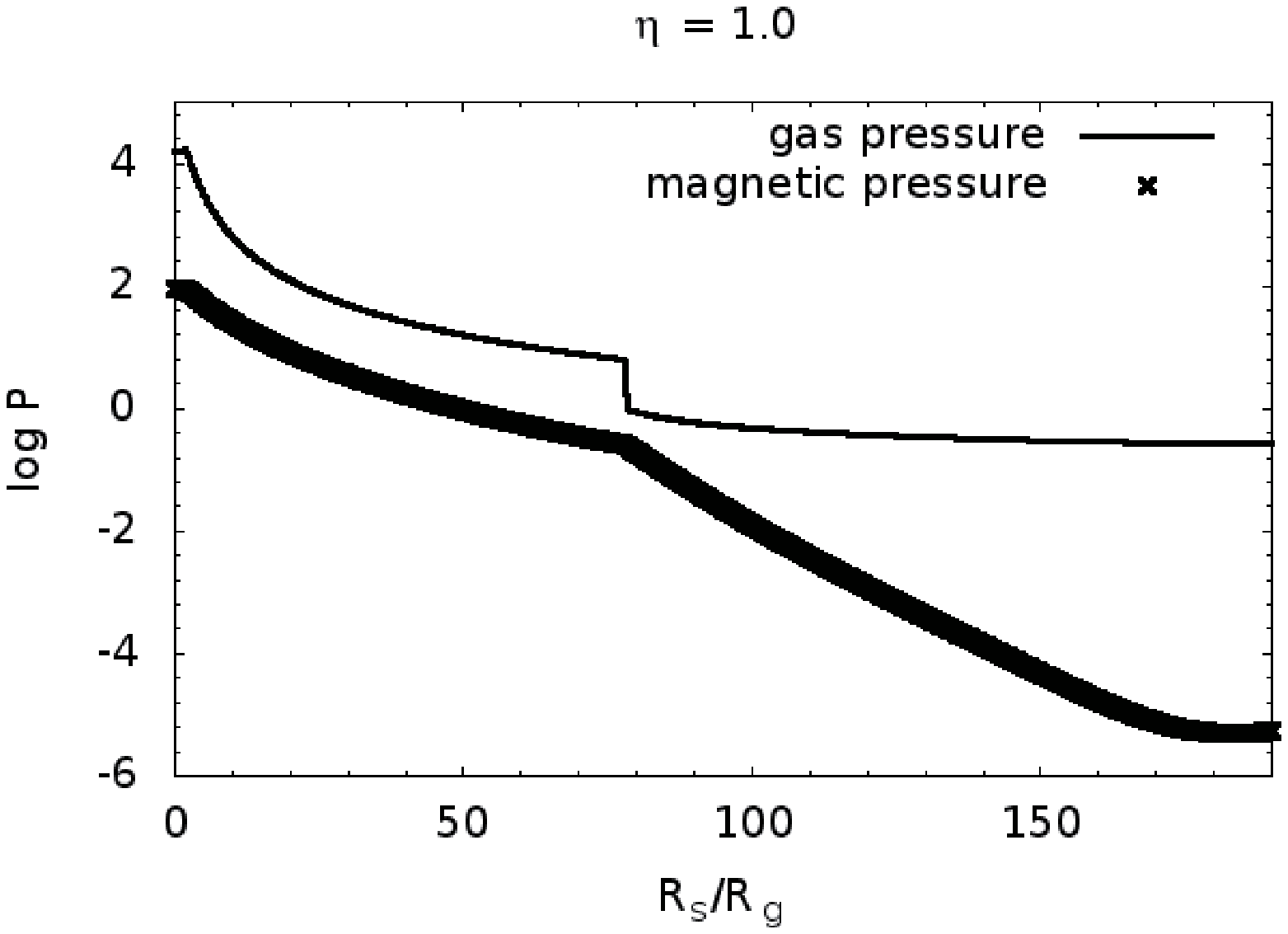}
        \label{fig:7d}
      \end{minipage}

  \begin{minipage}{0.5\linewidth}
        \centering
        \includegraphics[width=60mm,height=50mm]{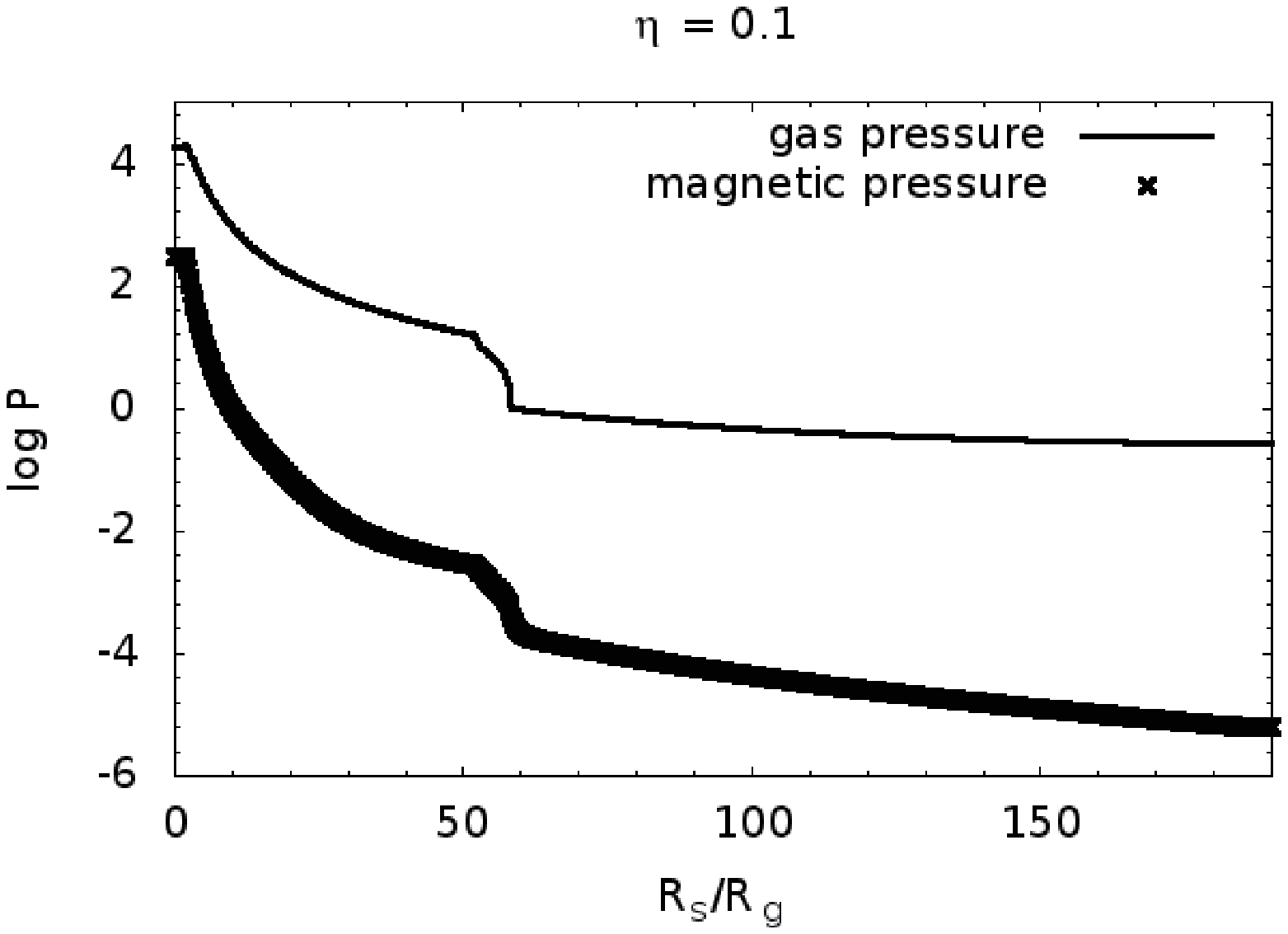}
        \label{fig:7c}
      \end{minipage}
    \end{tabular}
\caption{Radial profiles of gas pressure and magnetic pressure  which are space-averaged (between $-2R_{g}$ and $2R_{g}$ in z-direction) 
and time averaged (between $1.1 \times 10^{7}$ and $1.2 \times 10^{7}$ seconds) for resistive MHD flow with different resistivity, 
$\eta$ = $10^{-6}$, 0.01, 0.1 and 1 (in clockwise direction).}
  \label{Fig: fig7}
  \end{figure*}

\subsection{Effects of the resistivity on magnetized flow}

 The resistivity has  dissipative and diffusive characters in the magnetic field through the current density,
 similar to the viscosity in hydrodynamical flow and we expect the higher resistivity to  suppress the magnetic activity
 like magnetic turbulence.
 We examine the effects of resistivity through the time evolution of the magnetized flows with $\eta =
 10^{-6}$ to 1.0. In the case with the lowest resistivity $10^{-6}$, after a transient initial time evolution, 
 the magnetic field is amplified rapidly by the MRI and  the MHD turbulence develops  near the equatorial plane.
 Fig.~\ref{Fig: fig6} shows radial profiles of normalized Reynolds stress $\alpha_{gas}$ and normalized Maxwell stress $\alpha_{mag}$ 
for resistive MHD flow with different resistivity, $\eta$ = $10^{-6}$, 0.01, 0.1 and 1 (in clockwise direction).
These values are space-averaged (between $-2R_{g}$ and $2R_{g}$ in the z-direction) and the time averaged over the last
duration time.
Here we see that the maxwell stress $\Sigma_{mag}$ is larger by a factor of a few to ten 
 than the Reynolds stress $\Sigma_{gas}$ in cases of lower $\eta= 10^{-6}$ and 0.01 over the most region, 
 while in the higher $\eta$ the Reynolds stress mostly dominates over the Maxwell stress.   
 From this, we confirm that  the higher resistivity suppresses the Maxwell stress and then MHD turbulence,
 that is,  hydrodynamical mode dominates over magnetohydrodynamical mode.
As the result, in the case with the highest resistivity, the flow is dominated by the hydrodynamical quantities 
 at the outer radial boundary which are symmetric to the equator and the flow achieves a steady and symmetric state. 
Fig.~\ref{Fig: fig7} denotes radial profiles of the gas pressure
 and the magnetic pressure for MHD flow with resistivity, $\eta$ = $10^{-6}$, 0.01, 0.1 and 1
  (in clockwise direction). In all cases with different resistivity, the gas pressure dominates the magnetic pressure and the pressure
 distributions are not so different each other.

\begin{figure*}
    \begin{tabular}{cc}

      \begin{minipage}{0.5\linewidth}
        \centering
        \includegraphics[keepaspectratio, scale=0.4]{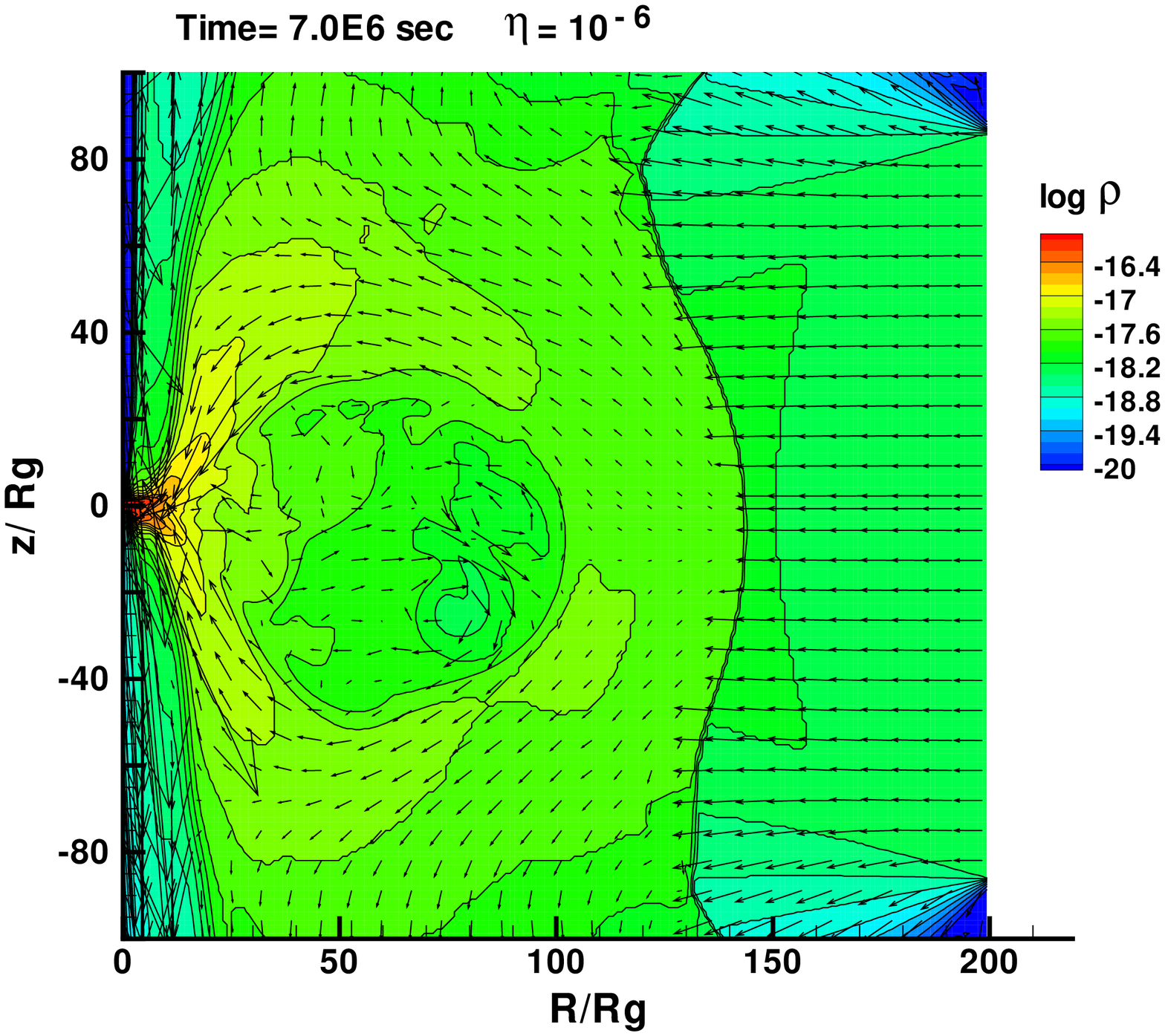}
        \label{fig:8a}
      \end{minipage}

      \begin{minipage}{0.5\linewidth}
        \centering
        \includegraphics[keepaspectratio, scale=0.4]{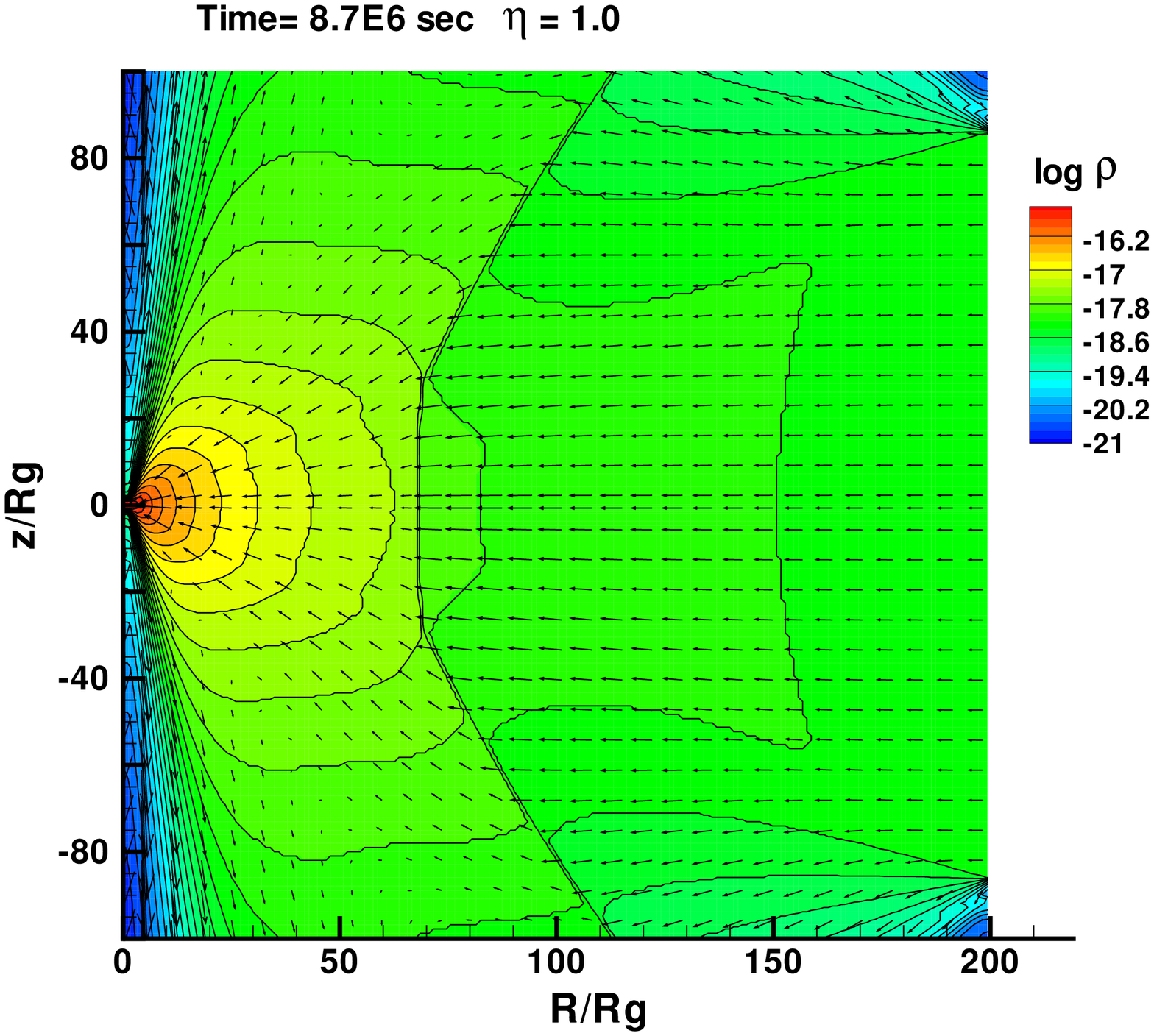}
        \label{fig:8b}
      \end{minipage} \\

    \end{tabular}
\caption {2D density contours and velocity vectors of  flows with $\eta = 10^{-6}$ and 1.0 at times $t$ = 7 $\times$ $10^{6}$ and 8.7 $\times$ $10^{6}$ seconds, 
respectively. In the former, variables of density and temperature are asymmetric to the equator and turbulent motions are
observed within the post-shock region  but in the latter case the flow is almost symmetric to the equator, and no turbulent motion appears. 
The shock locations are denoted by thick black contour lines.}
  \label{Fig: fig8}
  \end{figure*}

  Fig.~\ref{Fig: fig8} shows 2D density contours and velocity vectors at the later evolution of the flow with $\eta=10^{-6}$ and 1.0 
at times $t$ = 7 $\times$ $10^{6}$ and 8.7 $\times$ $10^{6}$ seconds respectively. 
Here, the location of  standing shock is distinguished  as the thick black contour lines, and the velocity vectors are 
 taken to be an arbitrary unit. In the low resistivity case, the density contours are asymmetric to the equator
  and turbulent motions are observed in the shocked region. While in the high resistivity case variables become symmetric to the equator and 
 no turbulent motion is observed. The flow features seem to return to the initial  hydrodynamical steady-state 
 but with a bit larger shock location $R_{s} \sim 70R_{g}$ than $R_{s} \sim 65 R_{g}$ in Fig.~\ref{Fig: fig2}, because the magnetic pressure 
contributes to the pressure balance to some extent in the shock location.

\subsection{Astrophysical significance}
\begin{figure*}
    \begin{tabular}{cc}

      \begin{minipage}{0.5\linewidth}
        \centering
        \includegraphics[keepaspectratio, scale=0.15]{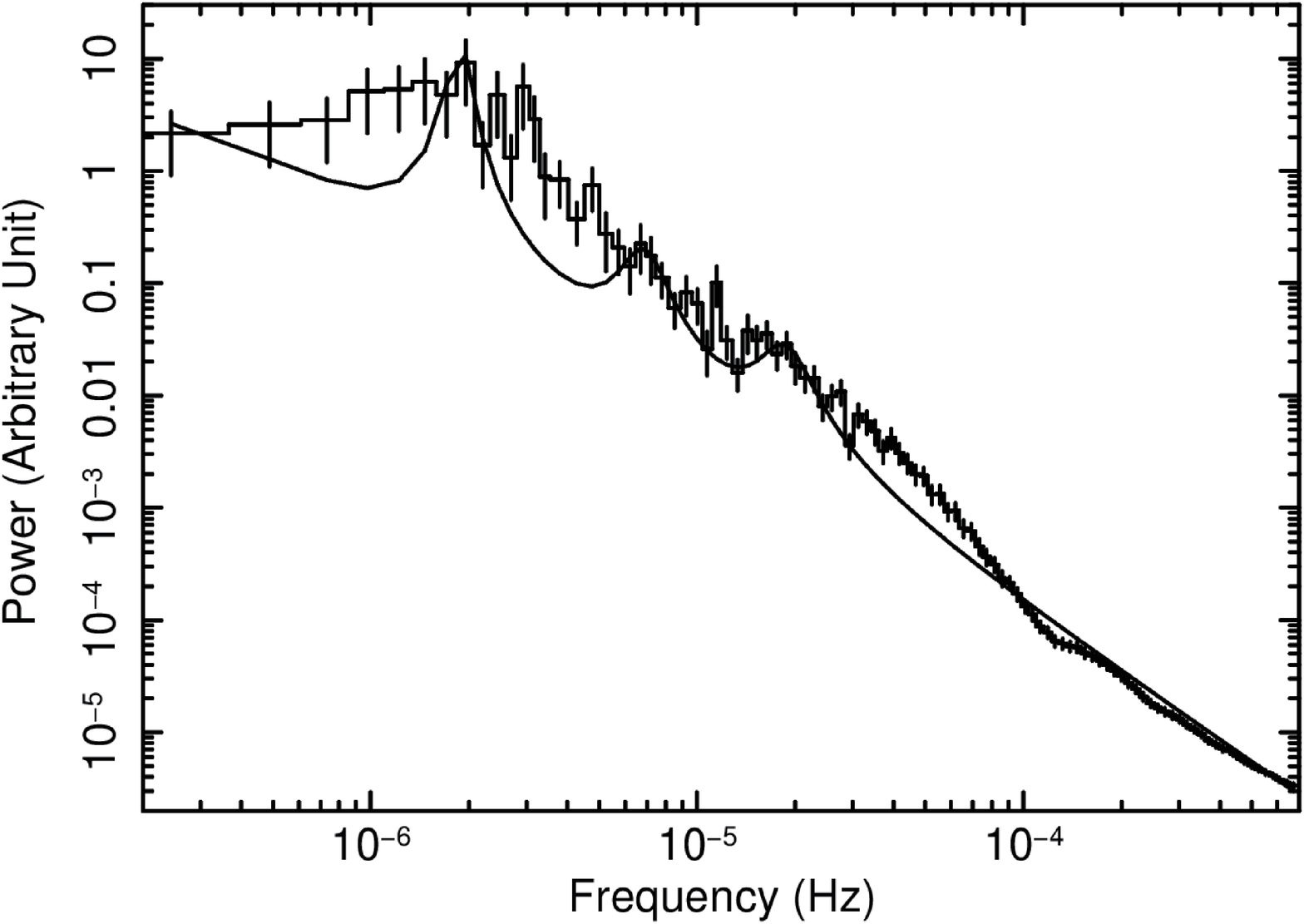}
        \label{fig:9a}
      \end{minipage}

      \begin{minipage}{0.5\linewidth}
        \centering
        \includegraphics[keepaspectratio, scale=0.15]{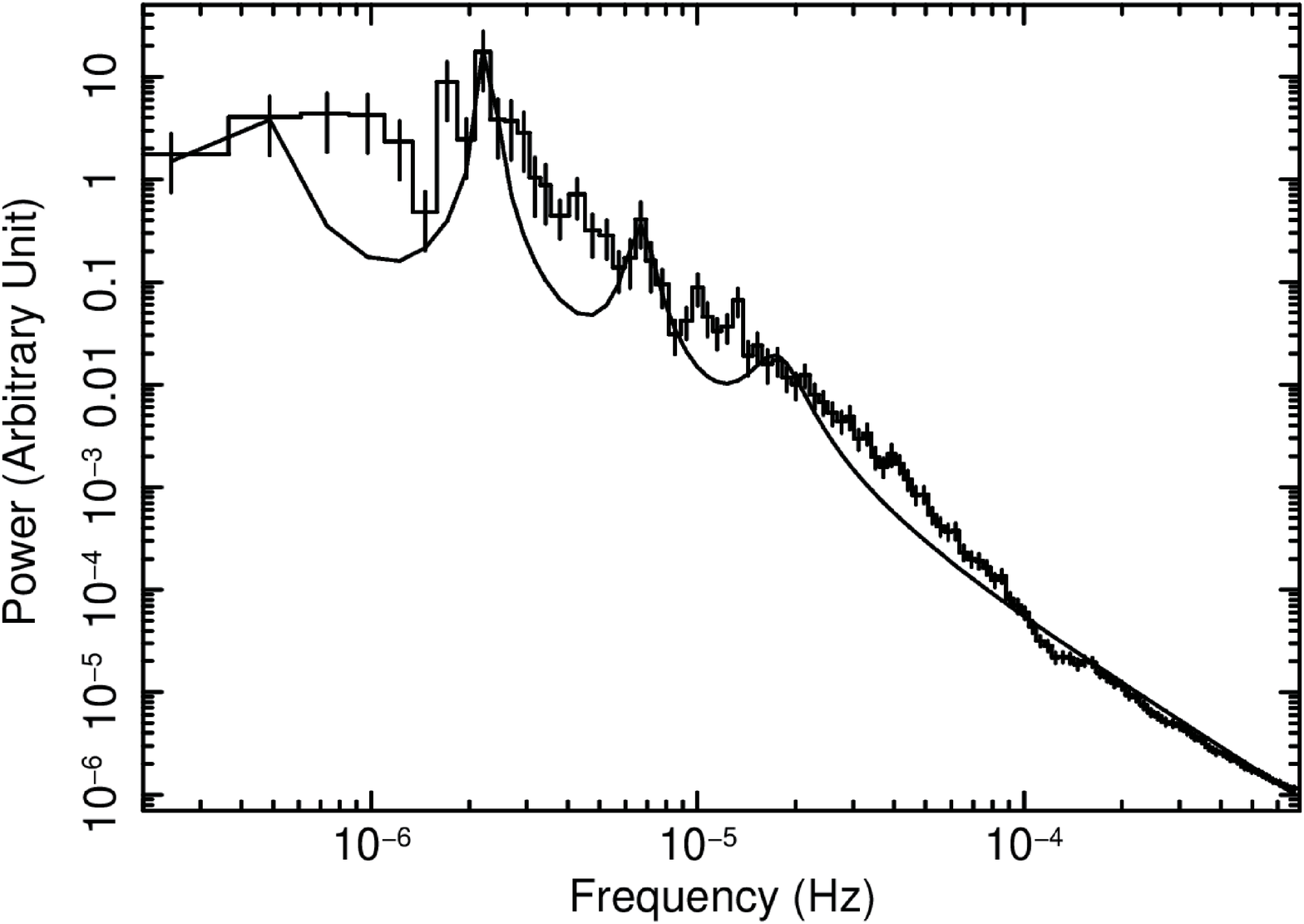}
        \label{fig:9b}
      \end{minipage} \\

  \begin{minipage}{0.5\linewidth}
        \centering
        \includegraphics[keepaspectratio, scale=0.15]{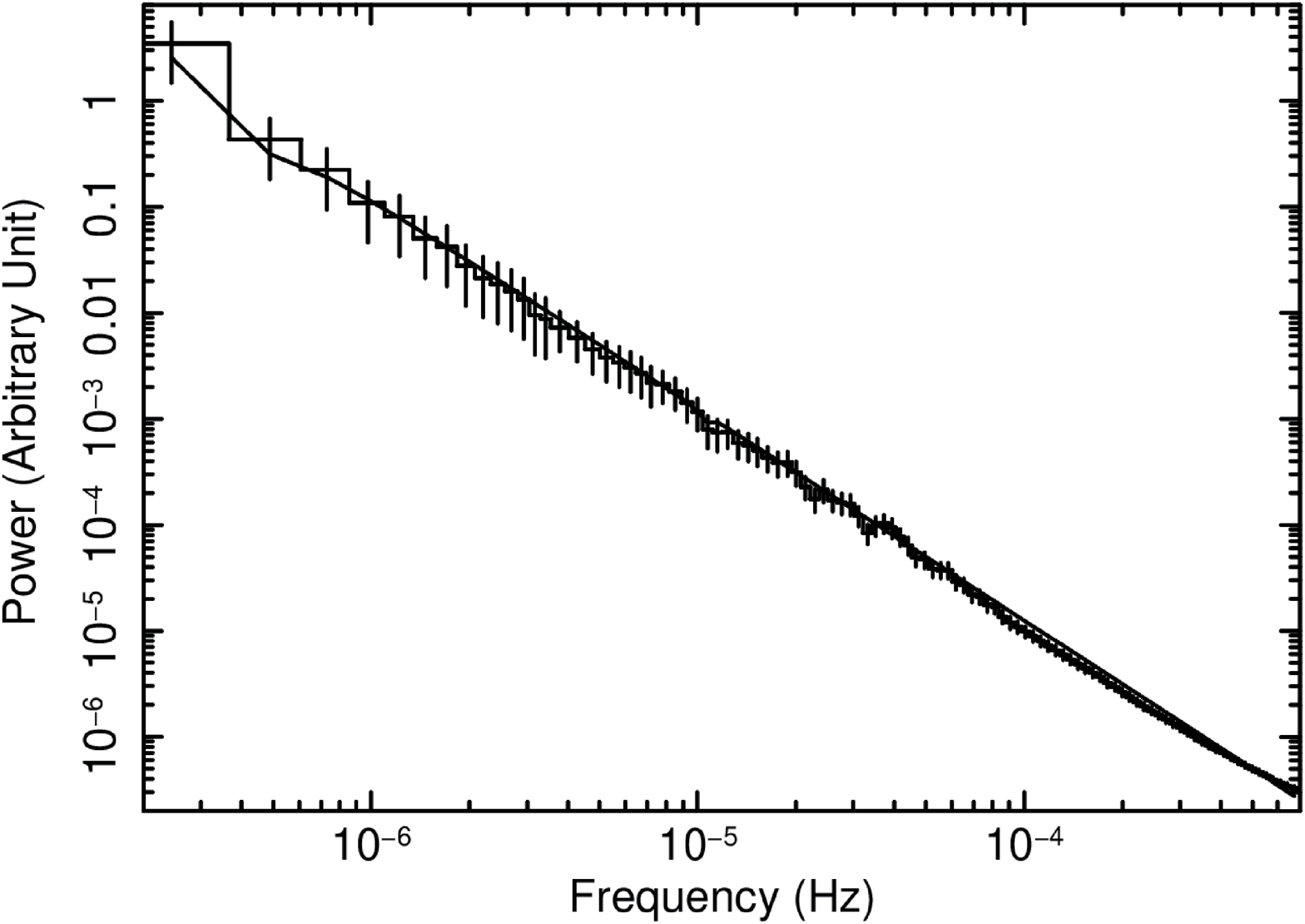}
        \label{fig:9d}
      \end{minipage}

  \begin{minipage}{0.5\linewidth}
        \centering
        \includegraphics[keepaspectratio, scale=0.15]{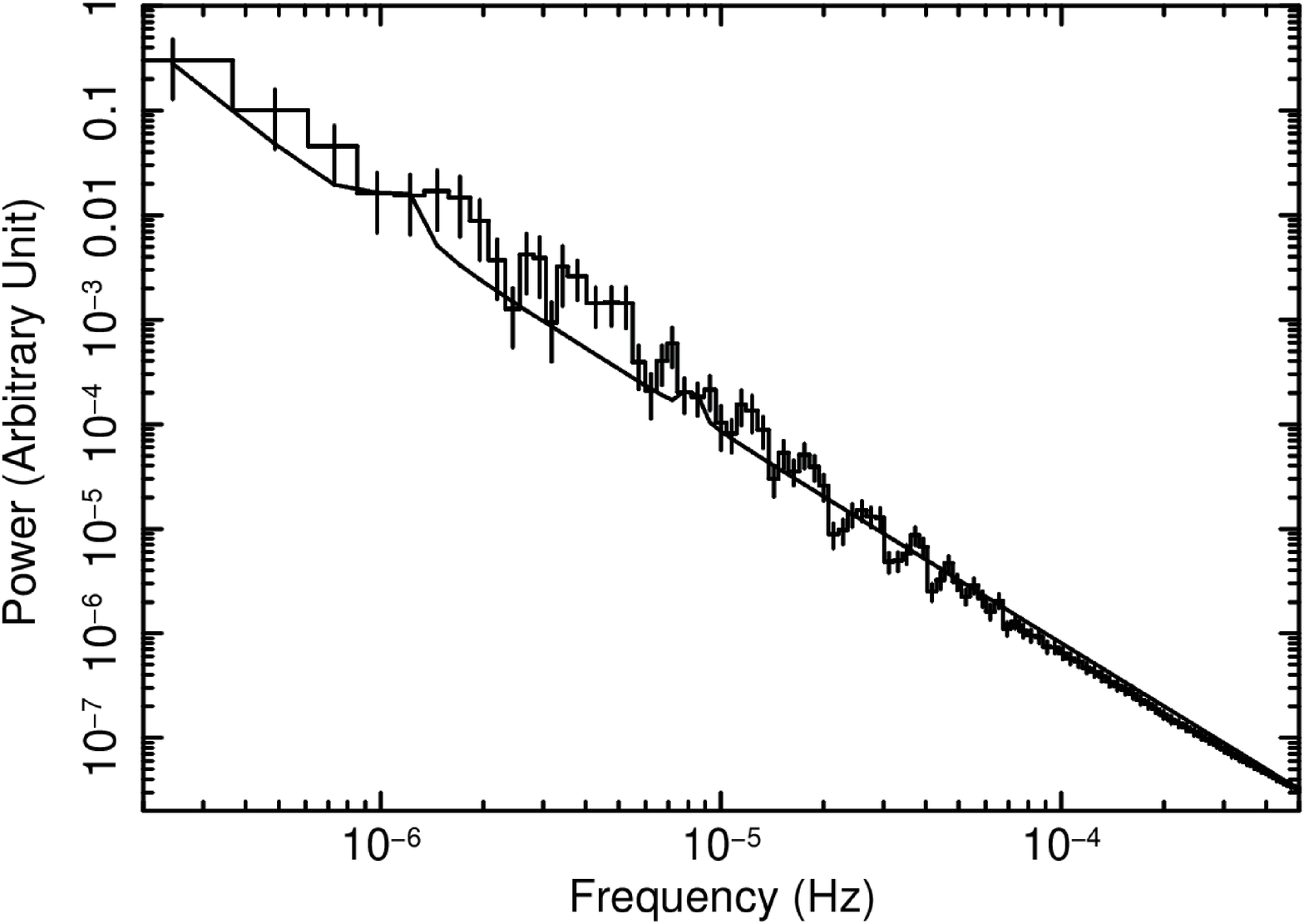}
        \label{fig:9c}
      \end{minipage}
    \end{tabular}
\caption{Power density spectra for resistive MHD flow with different resistivity values, $\eta$ = $10^{-6}$, 0.01, 0.1 and 1 (in clockwise direction).}
  \label{Fig: fig9}
  \end{figure*}

The present results for cases with low resistivity of $\eta= 10^{-6}$ and 0.01 are very similar to those for 
 the previous magnetized flow without resistivity (\cite{okuda19}).
 Adopting the same parameters of the flow and magnetic field as the present study,
 they found that the centrifugally supported shock moves back and forth between
  60 $R_{\rm g} \leq R \leq 170 R_{\rm g}$ and that another inner weak shock appears irregularly with rapid variations
 due to the interaction of the expanding high magnetic blob with the accreting matter below the outer shock.
 The process repeats irregularly with an approximate  time-scale of  
  (4 -- 5) $\times 10^5$ s ($\sim$ 5 days) with an accompanying smaller amplitude modulation with a period of
  $\sim 0.9\times 10^5$ s (25 hrs).
  In this respect, we also analyzed the time variability of the resistive magnetized flows.
 Fig.~\ref{Fig: fig9} show the power density spectra of luminosity for different values of resistivity.
 For $\eta$ = $10^{-6}$ and $0.01$, the peak (fundamental) frequency is estimated roughly to be at 2 $\times$ $10^{-6}$ along 
with two weak signatures (harmonics) at 7 $\times$ $10^{-6}$ and 2 $\times$ $10^{-5}$ Hz. 
These correspond to the periods of 5 $\times 10^5$s (5.8 days), 1.4 $\times 10^5$s (1.6 days), and 5$\times 10^4$s (0.6 day) , respectively 
and are comparable to two QPOs periods $\sim 5$ days and $\sim 1$ day  found in the non-resistive magnetized flow.
Therefore the QPO peak frequencies can be associated with periods of $\sim$ 5 -- 10 days and $\sim$ 1 day X-ray flares observed 
in the latest observations by Chandra, Swift, and XMM-Newton monitoring of Sgr A* (\cite{deg13, nei13, nei15, pon15}). 
On the other hand, for $\eta$ = $0.1$ and $1.0$, there is no clear peak frequency.
The average mass outflow rate $\sim 10^{-6}$ $M_{\odot} yr^{-1}$ and mass inflow rate $\sim 3 \times 10^{-6} M_{\odot} yr^{-1}$ obtained in 
small $\eta$ cases show a roughly good correspondence with the Chandra observations (\cite{wang13}) which suggest the presence of a high outflow rate 
that nearly balances the inflow rate.

\section{Summary}
We studied the effect of resistivity on standing shock in the magnetized flow around a black hole. 
The flow parameters of specific energy, $\epsilon$ = 1.98 $\times$ $10^{-6}$ and specific angular momentum, $\lambda = 1.35$, with 
$\Gamma$ = 1.6 have been considered to address the flow behaviour around Sgr A*. 
For flows with lower resistivity $\eta=10^{-6}$ and $0.01$, the luminosity and the shock location on the equator vary
quasi-periodically. These quasi-periodic oscillations  are attributed to the interactive result between the outer oscillating 
standing shock and the inner weak shocks occurring at the innermost hot blob.
The luminosity varies maximumly by a factor of ten around the average $L$ $\sim 3.0 \times 10^{34}$
  erg s$^{-1}$. The mass outflow  rate is very large as a few tens of percent of the input accretion rate.
 The MHD turbulence seems to play important roles in the outward transport of not only angular momentum but also
 accreting gas.
The power density spectra of luminosity variation show the peak frequencies which correspond to the periods of 
5 $\times 10^5$s (5.8 days),  1.4 $\times 10^5$s (1.6 days) and 5$\times 10^4$s (0.6 day) , respectively. 
 While for cases with higher resistivity $\eta=0.1$ and 1.0 the flow becomes steady and symmetric to the equator. 
Variable features of the luminosity disappear here and the steady standing shock is formed more outward compared with 
the hydrodynamical flow. The mass outflow rate is also high as $\sim$ a few tens $\%$ of the input gas.
The high resistivity considerably suppresses the magnetic activity such as the MHD turbulence and tends to form 
 the magnetized flow to be stable and symmetric to the equator.
The low angular momentum magnetized flow model with low resistivity has  possibility for the explanations of the high mass outflow rate $\sim$ 
as 10$\%$ of the Bondi accretion rate $\sim$ 1 $\times$ $10^{-5}$ $M_{\odot} yr^{-1}$ as suggested by Chandra observations (\cite{wang13}) and of the long-term 
flares with  $\sim$ one per day and $\sim$ 5 -- 10 days of Sgr A*  in the latest observations by Chandra, Swift, and XMM-Newton monitoring of Sgr A*.

\begin {acknowledgements}
CBS is supported by the National Natural Science Foundation of China under grant no. 12073021.
RA acknowledges support from National Science Foundation of China under grant No. 11373002, and 
Natural Science Foundation of Fujian Province of China under grant No. 2018J01007.

\end {acknowledgements}

\label{lastpage}

\end{document}